\documentclass[12pt,english]{article}
\pdfoutput=1

\usepackage[a4paper,lmargin=70pt,rmargin=70pt]{geometry}
\usepackage{setspace}
\usepackage{babel}
\usepackage{graphicx}
\usepackage{amsmath}
\usepackage{amssymb}
\usepackage[colorlinks=true,urlcolor=blue,anchorcolor=blue,citecolor=blue,filecolor=blue,linkcolor=blue,menucolor=blue,pagecolor=blue,linktocpage=true]{hyperref}

\usepackage[usenames, dvipsnames]{color}
\definecolor{orange}{RGB}{255, 140, 0}

\newcommand{\dho}{\partial}

\newcommand{\ed}{\,.}
\newcommand{\ec}{\,,}
\newcommand{\ecq}{\ec\quad}

\newcommand{\bC}{\ensuremath{\mathbb{C}}}

\newcommand{\cN}{\ensuremath{\mathcal{N}}}

\newcommand{\cZ}{\ensuremath{\mathcal{Z}}}

\newcommand{\cd}{\bar{c}}
\newcommand{\mymod}[2]{\ensuremath{#1 \;\mathrm{mod}\,\,#2}}
\newcommand{\NmodEight}{\ensuremath{N \;\mathrm{mod}\,\,8}}

\newcommand{\GUE}{\mathrm{GUE}}

\DeclareMathOperator{\trace}{Tr}

\newcommand{\be}{\begin{equation}}
\newcommand{\ee}{\end{equation}}
\newcommand{\tr}{\text{tr}\,}
\newcommand{\sgn}{\text{sgn}}

\begin{document}

\begin{flushleft} 
\today
\end{flushleft} 
\vspace{-1cm}
\begin{flushright} 
SU-ITP-16/19

YITP-16-124
\end{flushright} 

\vspace{0.1cm}

\begin{center}
  {\LARGE Black Holes and Random Matrices}
\end{center}
\vspace{0.1cm}
\begin{center}
	
  Jordan S. C{\sc otler}$^a$,
  Guy G{\sc ur-Ari}$^a$,  
  Masanori H{\sc anada}$^{abc}$,
  Joseph P{\sc olchinski}$^{de}$,

  Phil S{\sc aad}$^a$,
  Stephen  H. S{\sc henker}$^a$,
  Douglas S{\sc tanford}$^f$,

  Alexandre S{\sc treicher}$^{ad}$,
and  
  Masaki T{\sc ezuka}$^g$
      
\vspace{0.4cm}

$^a${\it Stanford Institute for Theoretical Physics,\\
Stanford University, Stanford, CA 94305, USA}

\vspace{0.2cm}

$^b${\it Yukawa Institute for Theoretical Physics, Kyoto University, Kyoto 606-8502, Japan}\\

\vspace{0.2cm}

$^c${\it The Hakubi Center for Advanced Research, Kyoto University, Kyoto 606-8501, Japan}\\

\vspace{0.2cm}

$^d${\it Department of Physics, University of California, Santa Barbara, CA 93106, USA}

\vspace{0.2cm}

$^e${\it Kavli Institute for Theoretical Physics,\\
University of California, Santa Barbara, CA 93106, USA}

\vspace{0.2cm}

$^f${\it Institute for Advanced Study, Princeton, NJ 08540, USA}

\vspace{0.2cm}

$^g${\it Department of Physics, Kyoto University, Kyoto 606-8502, Japan}\\

\end{center}

\onehalfspacing

\begin{center}
  {\bf Abstract}
\end{center}
We argue that the late time behavior  of horizon fluctuations in large anti-de Sitter (AdS) black holes is governed by the random matrix dynamics characteristic of quantum chaotic systems.   Our main tool is the Sachdev-Ye-Kitaev (SYK) model, which we use as a simple model of a  black hole. We use an analytically continued partition function $|Z(\beta +it)|^2$ as well as correlation functions as diagnostics. Using numerical techniques  we establish random matrix behavior at late times.  We determine the early time behavior exactly in a double scaling limit, giving us a plausible estimate for the crossover time to random matrix behavior.  
We use these ideas to formulate a conjecture about general large AdS black holes, like those dual to 4D super-Yang-Mills theory, giving a provisional estimate of the crossover time.  We make some preliminary comments about challenges to understanding the late time dynamics from a bulk point of view.

\newpage

\tableofcontents

\section{Introduction}

One of the deep questions in quantum gravity is the origin of the discrete spectrum of black hole microstates, from the {\it bulk} perspective of holographic duality.
For large black holes the $AdS$/CFT duality makes the answer clear from the boundary perspective --- a boundary field theory on a  compact space generically has a discrete spectrum of states.
But its origin from bulk gravity or string theory, even including nonperturbative effects like branes, is still mysterious.   

Maldacena \cite{Maldacena:2001kr} pointed out a signature of a discrete energy spectrum that can (in principle) be computed in the bulk --- the lack of decay of  two-point functions evaluated at very late time. Dyson, Lindesay, and Susskind \cite{Dyson:2002nt} applied these ideas to the study of correlators in de Sitter space.

To understand the way in which a two-point function diagnoses a discrete energy spectrum we can express it in the energy basis.
The two-point correlation function of a Hermitian operator\footnote{We assume that in a quantum field theory the operator is suitably smeared to eliminate any short distance divergences.} $O(t)$ at inverse temperature $\beta$ is given by 
\begin{align}\label{corr}
  G(t) &= 
  \frac{1}{Z(\beta)}
  \trace \left[ e^{- \beta H} O(t) O(0) \right] \cr
  &= \frac{1}{Z(\beta)} \sum_{m, n} e^{- \beta E_m} |\langle m | O | n \rangle |^2 e^{i(E_m -E_n)t} \ed
\end{align}
Here, $Z(\beta) = \trace \left( e^{-\beta H} \right)$ is the partition function and $\left| n \right>$ are energy eigenstates with energies $E_n$.
At early times we can replace the sum over eigenvalues by a coarse grained integral over a smooth density.  $G(t)$ will generically decay exponentially in time, but the decay does not continue indefinitely.
At late times the discreteness of the spectrum becomes important, and the phases in \eqref{corr} cause $G(t)$ to oscillate rapidly and erratically.  The correlation function is exponentially small and  no longer decays.

Holographically the coarse grained approximation is equivalent to a perturbative gravity calculation, and the exponential decay to quasinormal mode behavior \cite{Horowitz:1999jd}.  The decay  continues forever in this approximation.

There is a somewhat simpler diagnostic of a discrete energy spectrum, introduced in the black hole context by \cite{Papadodimas:2015xma}. We define
\begin{align} 
  Z(\beta,t) &\equiv \trace \left( e^{-\beta H - i H t} \right) 
  \ed \label{Z}
\end{align}
The quantity $Z(\beta,t)$ can be obtained by starting with $Z(\beta)$ and analytically continuing $\beta \to \beta + it$.  At late times $Z(\beta, t)$ also oscillates erratically.

The time average of an observable and its moments is a simple way to quantify its late time behavior.\footnote{The authors of \cite{Dyson:2002pf} use this idea in a closely related context.}
In fact, the time average of $Z(\beta,t)$ vanishes, which means that at late times this observable fluctuates around zero.
The typical size of the fluctuations can be studied by considering the squared quantity
\begin{align} \label{zz}
  \left| \frac{Z(\beta,t)}{Z(\beta)} \right|^2 &= \frac{1}{Z(\beta)^2} 
  \sum_{m, n} e^{- \beta (E_m+E_n)} e^{i(E_m -E_n)t}
  \ed
\end{align}
As in the case of the two-point function, the late time behavior of this quantity is generically complicated.
One can make progress by taking the long-time average,  where terms with oscillating phases average to zero and only terms with $E_m = E_n$ survive.
It is given by
\begin{align} \label{timeavgDeg}
  \lim_{T \to \infty} \frac{1}{T}\int_{0}^{T} dt
  \left| \frac{Z(\beta, t)}{Z(\beta)} \right|^2 
  &= 
  \frac{1}{Z(\beta)^2} \sum_{E} N_E^2 e^{-2\beta E} \ec
\end{align}
where $N_E$ is the degeneracy of the energy level $E$.
If the spectrum has no degeneracies ($N_E=1$), the long-time average becomes
\begin{align}
  \lim_{T \to \infty} \frac{1}{T}\int_{0}^{T} dt
  \left| \frac{Z(\beta, t)}{Z(\beta)} \right|^2 
  &= 
  \frac{Z(2 \beta)}{Z(\beta)^2} \ed \label{timeavg}
\end{align}
$Z$ generically scales as $e^{aS}$ where $S$ is the entropy  and $a$ is a positive constant. So \eqref{timeavg} scales as $e^{-aS}$.  In the holographic context   $S$ is the black hole entropy which scales as $1/g_s^2\sim 1/G_N$, where $g_s$ and $G_N$ are the string coupling and Newton constants of the bulk theory, so (\ref{timeavg}) is nonperturbative in the bulk coupling. For large black holes, $S$ is given by the thermal entropy of the boundary field theory, and it scales with the number of degrees of freedom.
In particular, we have $S \sim N^2$ in matrix theories like super-Yang-Mills (SYM) theory, and $S \sim N$ in vector theories like the Sachdev-Ye-Kitaev model \cite{Sachdev:1992fk,KitaevTalks}.
Either way, the quantity \eqref{timeavg} is non-perturbative in $1/N$.

Now, suppose we attempt to compute the left-hand side of \eqref{timeavg} by making a coarse grained approximation.  If we replace the discrete sum over states in \eqref{zz} by an integral over a smooth density we find that the long-time average vanishes.
  In holography, by analytically continuing saddle points we also find disagreement with \eqref{timeavg}.
  (See Section~\ref{N4} and also \cite{Dyer:2016pou}.)
Therefore, by studying how the long-time  decay of the partition function (or of the correlator) is avoided in gravity we are in fact probing the discreteness of the black hole spectrum --- a basic characteristic of its quantum nature.\footnote{It is sometimes said that this problem is related to the question of why a black hole has finite entropy. Indeed, in standard QM, finite entropy implies a discrete spectrum, but we note that  in disorder-averaged theories, or in a thermodynamic approximation, for example, one can effectively have a smooth but finite density of states.
}

From the bulk perspective, Maldacena initially suggested that an instanton might be responsible for the analogous $O(e^{-aS})$ root-mean-square (RMS) height of the correlator.   Barbon and Rabinovici \cite{Barbon:2003aq} pointed out that such an instanton might not describe the details of the irregular long-time fluctuations expected in the correlator.
Information loss in correlation functions was also studied in \cite{Fitzpatrick:2016ive,Fitzpatrick:2016mjq} in the context of $2d$ CFTs.
These questions have been difficult to address in standard holographic contexts like $\cN\!=\!4$ SYM, due to the difficulty in analyzing the chaotic boundary theory with sufficient precision.

The Sachdev-Ye-Kitaev (SYK) model \cite{Sachdev:1992fk,KitaevTalks} is a good laboratory to explore these questions. It is a quantum mechanical model of $N$ Majorana fermions with random $q$-fermion couplings that is soluble at large $N$.  
The theory is highly chaotic: at strong coupling it saturates \cite{kitaevfirsttalk,Polchinski:2016xgd,Maldacena:2016hyu} the chaos bound \cite{Maldacena:2015waa}, a property that is characteristic of black holes in Einstein gravity \cite{Shenker:2013pqa,kitaevfundamental,Shenker:2014cwa}. It realizes a (highly curved) description of a ``nearly $AdS_2$''/``nearly CFT$_1$'' system \cite{Almheiri:2014cka,Maldacena:2016hyu,Maldacena:2016upp,Jensen:2016pah,Engelsoy:2016xyb}.
As is the case for other vector models, there is an exact rewrite of the disorder-averaged model in terms of a functional integral over bilocal $O(N)$ singlet fields $G, \Sigma$ that presumably are related to the bulk description.\footnote{Higher dimensional versions of SYK have been constructed
 in \cite{Gu:2016oyy,Berkooz:2016cvq}. A supersymmetric generalization of the model has been constructed in \cite{Fu:2016vas}. A multiflavor version has been constructed in \cite{Gross:2016kjj}.  Other related work includes \cite{Jevicki:2016bwu,Almheiri:2016fws,Bagrets:2016cdf,Cvetic:2016eiv}.}

The SYK model has several other properties that make it useful in the study of late time properties.   The average over the random couplings should rattle
the energy eigenvalues sufficiently to make the rapidly oscillating terms in equations \eqref{corr},\eqref{zz} average to zero at a {\it fixed} time, making these quantities smooth functions of time.  This makes them more amenable to study.  In addition, the model is computationally simple enough that numerical methods can yield significant insight \cite{Fu:2016yrv,You:2016ldz}.  (After we had finished our numerical analysis  the paper \cite{Garcia-Garcia:2016mno} appeared. It has significant overlap with our numerical
results.)

One  goal of this paper is to explore the late time behavior of the SYK model.  We present numerous numerical results about such behavior in the model, and interpret them using a variety of analytic and conceptual arguments.  One of our key findings is a close relationship between the late time behavior of the model and the behavior of random matrices.\footnote{Another discussion of random matrices in black hole physics is \cite{Balasubramanian:2014gla}. Recent discussion of a connection between chaotic systems and random ensembles, including observables generalizing $\langle |Z(\beta,t)|^2\rangle$, appears in \cite{Roberts:2016hpo}.} 

It is a widely held conjecture \cite{mehta2004random} that the spacing statistics of nearby energy levels in quantum chaotic systems should be well approximated by an appropriate random matrix ensemble.   Since late times corresponds to small energy differences our result is a natural one.

Building on these observations we can make a plausible conjecture about the behavior of more complicated holographic systems, like the Type IIB $AdS_5$ / $\cN=4$ SYM system.

\subsection{Summary of results}

Here we give an outline of the paper and summarize the main results.
In Section~\ref{model} we introduce the SYK model.
Then in Section~\ref{spectral} we write down the spectral form factor, which is
given by $|Z(\beta,t)|^2/Z(\beta)^2$ averaged over the random couplings.
At late times this quantity goes over to a plateau value given approximately by \eqref{timeavg}, which characterizes the discreteness of the spectrum.
By numerically computing this quantity we find that its late time behavior exhibits an interesting feature, see Figure~\ref{fig:g-SYK}.
Starting at $t=0$, the spectral form factor first dips below its plateau value
and then climbs back up in a linear fashion (we call this region the `ramp'), joining onto the plateau.
This behavior is readily explained if we approximate the SYK Hamiltonian by a Gaussian random matrix, as shown in Figure~\ref{fig:g-RMT}.
Further evidence for the relation between the late time behavior and random
matrix theory (RMT) is given in Section~\ref{nn}, where we show the relation
between the choice of RMT ensemble (GUE, GOE, or GSE) and the detailed shape of
the late time behavior in SYK.
See Figure~\ref{fig:g-majorana-beta0}.

In Section~\ref{thermo} we make a digression to discuss the thermodynamic properties of SYK.
We compute the entropy and energy numerically, and by
extrapolating these results to infinite $N$ we find excellent agreement with existing analytical calculations carried out in the large $N$ limit.
This serves as an incisive check both on our results and on existing analytic calculations.

In Section~\ref{rampRMT} we review the analytical origin of the ramp  and plateau in RMT, and the relation of the ramp to the phenomenon of spectral rigidity.
We show that the ramp can be understood as a perturbative effect in RMT (though not as a perturbative $1/N$ effect in SYK, as we explain).

In Section~\ref{rampSYK} we explain the early-time power-law decay in SYK visible in Figure~\ref{fig:g-SYK}. This is related to the low energy portion of the spectrum, dominant in the large $N$, large $\beta J$ limit, that is described by the Schwarzian theory of reparametrizations.  We argue this is exact in  a double scaling limit.   In the large $N$, large $\beta J$ limit  a sector of the model \cite{Maldacena:2016hyu, Maldacena:2016upp} is dual to a dilaton gravity \cite{Teitelboim:1983ux,Jackiw:1984je} black hole in $AdS_2$ .  We argue that the subsequent linearly growing ramp  and the plateau should survive in this limit,
 suggesting a connection between the late time behavior of black
holes and random matrix theory.

In Section~\ref{sec:corr} we discuss a similar ramp that appears in SYK correlators. We work out the conditions under which the fermion two-point function exhibits the ramp/plateau structure of the spectral form factor, and check these results numerically.

In Section~\ref{single} we consider the behavior of the spectral form factor for
a single realization of the random couplings.
The motivation here is to make contact with theories such as Yang-Mills which do
not involve an averaging over couplings.
For a single realization the spectral form factor exhibits large fluctuations
even at large $N$, but we argue that by time averaging (and no disorder
averaging) the underlying ramp/plateau structure can be brought into view.

In Section~\ref{N4} we make a connection with $\cN=4$ SYM, giving a preliminary estimate of the gravity saddle points that give the early-time decay of $|Z(\beta,t)|^2$. We also argue that there should be a subsequent long period of time where this quantity is growing and dominated by `ramp' physics, folded against the coarse-grained density of states of the SYM theory.

We conclude and discuss future directions and ongoing work in
Section~\ref{discussion}.

Several appendices contain additional results and discussion.

In Appendix~\ref{P} we review the particle-hole symmetry of the SYK model, whose properties depend on $N~\mathrm{mod}~8$ \cite{PhysRevB.83.075103,You:2016ldz}.

In Appendix~\ref{doubleScaledSYK} we discuss the double-scaled limit of SYK, where the disorder-averaged density of states can be computed exactly.

In Appendix~\ref{toyapp} we consider a toy model of the $G,\Sigma$ path integral, which is an exact rewrite of the SYK model in terms of bosonic bilocal fields.
We explain how the original fermionic behavior can arise from these bosonic variables.

In Appendix~\ref{subleadingsaddles} we again consider the $G,\Sigma$ formulation of the model.
We point out the existence of a family of subleading saddle points that show up both in the SYK model and in the integrable version $q=2$ of it.
We explain why this infinite family of saddle points does not significantly affect the thermodynamics of the model at large $N$. 

In Appendix~\ref{qequal2} we further discuss these saddle points in the integrable $q=2$ version of the SYK model, and show how they lead to a simple kind of random matrix theory behavior at late times. 

In Appendix~\ref{Nq} we make some preliminary remarks on the origin of the amplitude of the ramp in SYK.

In Appendix~\ref{Gramp} we present constraints on a simple single saddle point explanation of the ramp in SYK correlators.

Finally, in Appendix~\ref{dataapp} we present additional numerical data.

\section{The Sachdev-Ye-Kitaev model}
\label{model}

Consider $N$ Majorana fermions $\psi_a$ ($a=1,\dots,N$) in 0+1 dimensions that obey the algebra $\{ \psi_a, \psi_b \} = \delta_{ab}$.
The Hamiltonian is\footnote{
We follow the conventions of \cite{Maldacena:2016hyu} and specialize to $q=4$, where $q$ is the number of fermions interacting in each term of the Hamiltonian.
}
\begin{align}
  H = \frac{1}{4!} \sum_{a,b,c,d} J_{abcd} \psi_a \psi_b \psi_c \psi_d 
  = \sum_{a<b<c<d} J_{abcd} \psi_a \psi_b \psi_c \psi_d 
  \ed \label{Hmajorana}
\end{align}
The coupling tensor $J_{abcd}$ is completely anti-symmetric, and each independent element is a random real number chosen from a Gaussian distribution with zero mean and variance given by $\sigma^2 = \frac{3!}{N^3} J^2$.
The Hilbert space has dimension
\begin{equation}
L\equiv \text{dim. of Hilbert space} = 2^{N/2},
\end{equation}
and we set $J=1$ for convenience.

In this work we mainly focus on the model with 4-fermion interactions, although we will sometimes discuss the generalization where the fermions interact in groups of $q$.

For $N$ even it is often useful to implement the model using $N_d = \frac{N}{2}$ Dirac fermions $c_i$ ($i=1,\dots,N_d$) by defining 
\begin{align}
  \psi_{2i} &= \frac{c_i + \cd_i}{\sqrt{2}} \ecq
  \psi_{2i-1} = \frac{i ( c_i - \cd_i )}{\sqrt{2}} \ed
  \label{chiDirac}
\end{align}
The Dirac fermions satisfy the algebra
\begin{align}
  \{ c_i, \cd_j \} = \delta_{ij} \ecq
  \{ c_i, c_j \} = 0 \ecq
  \{ \cd_i, \cd_j \} = 0 \ed
\end{align}
We can write down a fermion number charge given by $Q = \sum_{i=1}^{N_d} \bar{c}_i c_i$.
The Hamiltonian \eqref{Hmajorana} does not preserve this charge, but it does preserve charge parity $(\mymod{Q}{2})$.
Therefore, the Hamiltonian has two blocks corresponding to even and odd values of $Q$.

\section{Spectral form factor}
\label{spectral}

We define disorder-averaged analogs of the quantity in equation \eqref{zz} as follows.
\begin{align}
  g(t;\beta) &\equiv \frac{\langle Z(\beta,t) Z^*(\beta,t) \rangle_J}
  {\langle Z(\beta) \rangle_J^2} \ec \label{g}
  \\
  g_{d}(t;\beta) &\equiv \frac{\langle Z(\beta,t) \rangle_J \cdot \langle Z^*(\beta,t) \rangle_J}
  {\langle Z(\beta) \rangle_J^2} \ec \label{gd}
  \\
  g_{c}(t;\beta) &\equiv g(t;\beta) - g_{d}(t;\beta) \ed \label{gc}
\end{align}
Here $\langle \cdot \rangle_J$ denotes the disorder average --- the average over the ensemble of random couplings.
$Z(\beta,t)$ was defined in \eqref{Z}.
As discussed in the introduction, the late-time behavior of these quantities probes the discreteness of the spectrum, similar to the late-time behavior of two-point functions.
Notice that we are working with annealed quantities, meaning that we are taking the disorder average separately in the numerator and denominator.
This is in contrast with quenched quantities such as $\left< |Z(\beta,t)|^2 / Z(\beta)^2 \right>_J$.
The advantage of working with annealed quantities is that they require a finite number of replicas in analytic calculations ($g$ requires two replicas, $g_d$ requires just one), whereas quenched quantities require an  arbitrary number of replicas.\footnote{ Numerically, we find that the quenched and annealed versions of $g(t;\beta)$ remain well within a percent of each other for all times and values of $\beta$ we considered, and the difference appears to decrease with $N$.  (At infinite temperature the annealed and quenched quantities are in fact equal because $Z(\beta=0) = \trace(1)$ is independent of the random couplings.)}

\begin{figure}[h]
  \centering
  \includegraphics[width=0.6\textwidth]{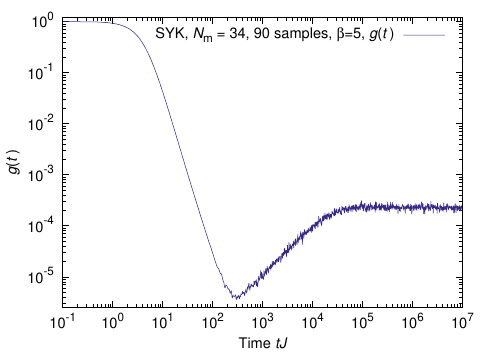}
  \caption{
A log-log plot of SYK $g(t; \beta=5)$, plotted against time for $N=34$.
Here we use the dimensionless combination $tJ$ for time.
Initially the value drops quickly, through a region we call the \textit{slope}, to a minimum, which we call the \textit{dip}.
After that the value increases roughly linearly, $\sim t$, until it smoothly connects to a plateau around $tJ = 3\times 10^4$.
We call this increase the \textit{ramp}, and the time at which the extrapolated linear fit of the ramp in the log-log plot crosses the fitted plateau level the \textit{plateau time}.
The data was taken using $90$ independent samples, and the disorder average was taken for the numerator and denominator separately.
}
  \label{fig:g-SYK}
\end{figure}

Now we present one of the central results of this work, $g(t)$ for the SYK model.   
In Figure~\ref{fig:g-SYK} we present $g(t;\beta=5)$ for $N=34$, computed numerically.\footnote{All numerical results in this paper were computed by fully diagonalizing the SYK Hamiltonian for independently generated Gaussian random couplings, computing the relevant quantity, and then taking the mean.}
Notice that $g(t)$ at early times does not simply join onto the late-time plateau, but instead dips below the plateau and then climbs back up.
One goal of this work is to understand the source and implications of this behavior, and to estimate how prevalent it is both in SYK (for various values of the parameter $\beta J$) and in quantum field theory in general.

Notice that $g(t)$ is smooth, and does not exhibit the large fluctuations that one expects at late times in a typical quantum theory.
This is due to the disorder average, which smooths out the fluctuations exhibited by each realization of the random couplings. 
(Some fluctuations are apparent at late times, but these are an artifact due to the finite number of samples used in the computation.
We will discuss this point further in Section~\ref{single}.)

We will be discussing the curve $g(t)$ at length, so let us point out the main features in this plot and introduce some nomenclature.
Starting with $t=0$, at early times the value of $g(t)$ drops quickly along what we will call the `slope', until it reaches a minimum at the `dip~time' $t_d$.
Next comes a period of linear growth that we will call the `ramp'.
It ends at the plateau time $t_p$, and beyond this we have an almost constant value of $g(t)$ that we call the `plateau'.
The plateau height is equal to the long-time average of $g(t)$.
On the plateau only the $E_n = E_m$ terms in the sum \eqref{zz} survive, and the height of the plateau is $2Z(2\beta)/Z^2(\beta) \sim e^{-a S}$, in accordance with \eqref{timeavgDeg}.
The factor of 2 is due to a 2-fold degeneracy in the spectrum (see Appendix~\ref{P}).

Quantities such as $g$, $g_d$, and $g_c$ are studied extensively in the field of quantum chaos. 
In particular, $g(t)$ (typically used with $\beta=0$) is called the \textit{spectral form factor} and it is a standard diagnostic of the pair correlation function of energy eigenvalues. We will often refer to $g(t)$ by this name.
It supplies information about the correlations of eigenvalues at different energy separations.\footnote{ 
The spectral form factor contains information about the pair correlation between well-separated eigenvalues that the (perhaps more familiar) diagnostic of the nearest-neighbor energy spacing distribution does not.
Conversely, the nearest-neighbor level spacing distribution contains information about multi-point correlation functions of nearby eigenvalues that the spectral form factor does not.}   

One of the basic conjectures in the field of quantum chaos is that the fine grained energy eigenvalue structure of a chaotic system is the same as that of a random matrix chosen from one of the standard Dyson ensembles \cite{Dyson:1962es}: Gaussian Unitary Ensemble (GUE), Gaussian Orthogonal Ensemble (GOE), or Gaussian Symplectic Ensemble (GSE). (For reviews, see \cite{mehta2004random,Guhr:1997ve}.)
The particular ensemble to use depends on the symmetries of the original Hamiltonian.
Random matrix theory can then be used to compute certain quantities (such as the spectral form factor) that are sensitive to eigenvalue correlations.
You, Ludwig and Xu \cite{You:2016ldz} first discussed the quantum chaotic properties of SYK by studying the distribution of spacings between nearest-neighbor energy levels, another standard quantum chaos observable.
They showed that the distribution is consistent with RMT predictions.

\begin{figure}[h]
  \centering
  \includegraphics[width=0.6\textwidth]{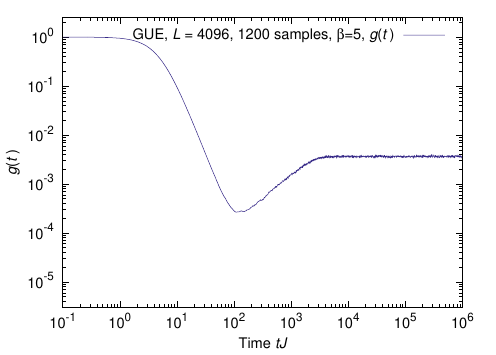}
  \caption{A log-log plot of $g(t; \beta=5)$ against time for GUE random matrices, dimension $L=2^{12}$.
	A dip, ramp and plateau structure similar to  Fig.~\ref{fig:g-SYK} is apparent.}
  \label{fig:g-RMT}
\end{figure}

In Figure~\ref{fig:g-RMT} we present $g(t;\beta=5)$ for the GUE ensemble of matrices of rank $L_{\rm RMT}=2^{12}$, computed numerically, with a normalization such that the eigenvalues typically lie in the range $-2<\lambda<2$ (see \eqref{ZGUE}).  At $\beta = 0$ the height of the plateau is of order $1/L_{\rm RMT}$ and the plateau time is at $t$ of order $L_{\rm RMT}$, the inverse mean level spacing. 

Note the similarity between the RMT result and the SYK result, and in particular the presence of the ramp and the plateau.
We will argue that the late-time behavior of the spectral form factor in SYK can be explained by random matrix theory. The early time behavior of RMT differs from SYK, although it is not obvious from the plots. The typical eigenvalue density has different dependence on energy in the two systems, which leads to somewhat different initial decays. Moreover, at early times RMT is governed by a perturbative expansion in $1/L$, while SYK is governed by an expansion in $1/N$. On the other hand, at times well beyond the dip, $g(t)$ is determined by eigenvalue correlations on scales much smaller than the total width of the spectrum, and there one expects to find agreement between SYK and RMT.

What is the physical origin of the ramp in RMT?
Eigenvalues of generic matrices repel, so near degeneracies are extremely unlikely.  This causes the plateau.  The time of onset of the plateau is determined by the scale of near neighbor eigenvalue spacings.   The ramp, though,  is due to the repulsion between eigenvalues that are far apart in the spectrum.   This repulsion, when balanced against the effects that keep the energy finite, gives rise to a very rigid eigenvalue structure.   This phenomenon is referred to as long-range spectral rigidity \cite{Dyson:1962es,mehta2004random,Guhr:1997ve}.   More quantitatively, if $\delta E_n$ denotes the deviation of an energy from its average value, then at leading order $\langle \delta E_n \delta E_m \rangle \sim \log |n-m| $.    For comparison, if the eigenvalues formed a one dimensional crystal with harmonic near neighbor interactions, then  $\langle \delta E_n \delta E_m \rangle \sim |n-m| $, a much less rigid behavior \cite{Dyson:1962es,mehta2004random,Guhr:1997ve}.  The $ \log |n-m| $ form, after suitable processing we will discuss below, accounts for the linear behavior of the ramp.  The ramp lies below the plateau because repulsion causes the eigenvalues to be anticorrelated.

\subsection{The ramp and the eightfold way}
\label{nn}

We now present further evidence of the relation between random matrix theory and the presence of the ramp in the SYK spectral form factor.

The Hamiltonian of a chaotic theory is generally believed to resemble a random matrix when studied at sufficiently fine energy resolution.
One basic property of random matrices is their nearest-neighbor level statistics, namely the distribution of the distance $s$ between pairs of neighboring energy levels \cite{wigner1956proceedings}.\footnote{
More precisely, one considers the distribution of spacings between \textit{unfolded} energy levels \cite{bohigas1984chaotic}.
These are the levels one obtains by making a change of variables such that the mean level spacing becomes one everywhere.
For further details, see \cite{Guhr:1997ve}.
}
The nearest-neighbor statistics of an integrable theory follow an exponential distribution $e^{-s}$, while those of a chaotic theory generally follow one of the three reference ensembles GUE, GOE, and GSE.
The particular ensemble depends on the symmetries of the Hamiltonian.

You, Ludwig and Xu \cite{You:2016ldz} studied the nearest-neighbor level spacing distribution in SYK.
They made the important point that all three Gaussian RMT ensembles are implemented in the model as we now review.

\begin{figure}[h]
  \centering
  \includegraphics[width=0.6\textwidth]{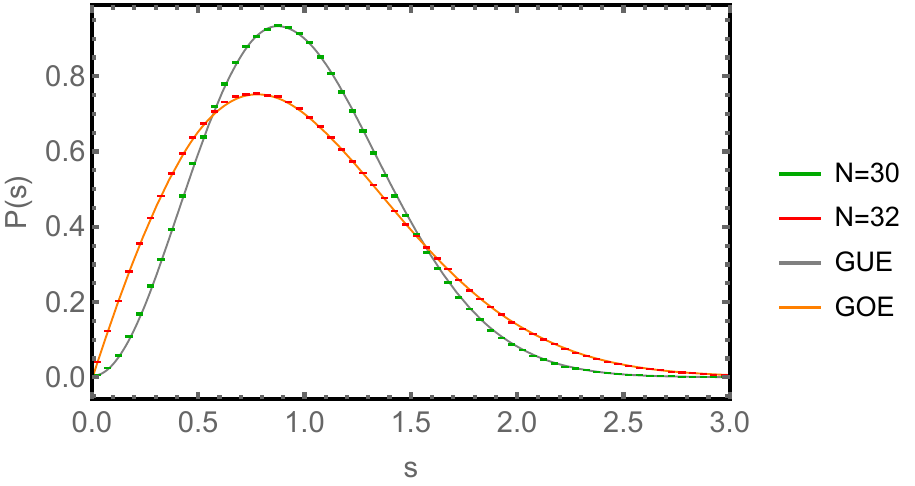}
  \caption{Unfolded nearest-neighbor level spacing distribution for SYK vs. RMT. Here $s$ is measured in units of the mean spacing. Semi-analytical exact large $L$ results (correcting the Wigner surmise) for the RMT $P(s)$ are available \cite{mehta1960statistical,gaudin1961loi}, but we computed the RMT curves from $L = 12870$ exact diagonalization data.}
  \label{fig:maj-nearest-neighbor-with-exact-rmt-0-4.pdf}
\end{figure}

The SYK model has a particle-hole symmetry given by \cite{PhysRevB.83.075103,You:2016ldz,Fu:2016yrv}
\begin{align}
  P = K \prod_{i=1}^{N_d} (\cd_i + c_i) \ec
\end{align}
where $K$ is an anti-linear operator.
The properties of this operator determine the class of RMT statistics of each charge parity sector of the Hamiltonian.
In particular, the statistics are determined by the value of $(\NmodEight)$ as follows (see Appendix~\ref{P} for details).
\begin{itemize}
  \item When $\NmodEight = 2$ or 6, the symmetry $P$ maps the even and odd parity sectors to each other. Individual sectors do not have any anti-linear symmetry, and the corresponding ensemble of each sector is GUE.
  \item When $\NmodEight = 0$, $P$ maps each sector to itself and $P^2=1$.
    The corresponding ensemble is GOE.
  \item When $\NmodEight = 4$, $P$ again maps each sector to itself but now $P^2=-1$.
    The corresponding ensemble is GSE.
\end{itemize}
Figure~\ref{fig:maj-nearest-neighbor-with-exact-rmt-0-4.pdf} shows the nearest-neighbor statistics of SYK with $N=30,32$, and we see excellent agreement with RMT predictions.

While the nearest-neighbor spacing distribution is sensitive to correlations between adjacent energy levels, the spectral form factor probes correlations between energy levels at larger separations.
The $t$ parameter in $g(t)$ determines the scale of the energy differences being probed.
As discussed above, beyond the plateau time only individual energy levels are probed, while at earlier times (and in particular on the ramp) $g(t)$ is sensitive to correlations between levels that are much farther apart than the mean level spacing.
The structure of these correlations depends on the ensemble.
The three RMT ensembles all exhibit a ramp and a plateau but with slightly different shapes: In GUE the (unfolded) ramp and plateau connect at a sharp corner, in GOE they connect smoothly, and in GSE they connect at a kink.\footnote{
See, for instance, Figure 10 in \cite{Guhr:1997ve}.
}

\begin{figure}[h]
  \centering
  \includegraphics[width=0.48\textwidth]{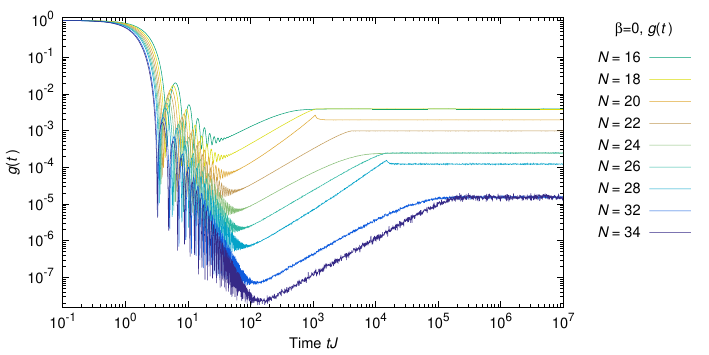}
  \includegraphics[width=0.48\textwidth]{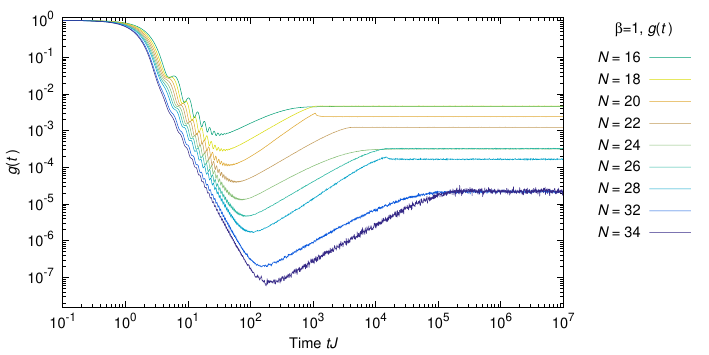}
  \includegraphics[width=0.48\textwidth]{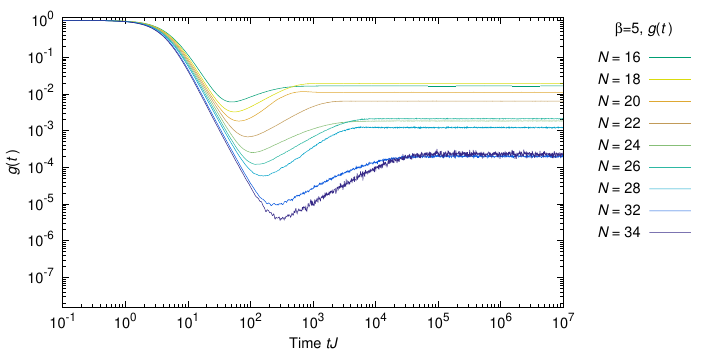}
  \caption{
SYK $g(t, \beta)$ with $\beta=0,1,5$ and various $N$ values.
The value at late times, which is equal to plateau height $g_p$, matches with $N_E Z(2\beta)/Z(\beta)^2$ as discussed in Appendix~\ref{App:dtime}.
Here $N_E$ is the eigenvalue degeneracy, $2$ for $(N~\mathrm{mod}~8)\neq 0$ and $1$ for $(N~\mathrm{mod}~8)= 0$.
As explained in the main text, the shape of the ramp and the plateau depends on the symmetry class, and the agreements with the counterparts in the RMT with GUE, GOE, and GSE are good.
The numbers of samples are 1 200 000 ($N=16$), 600 000 ($N=18$), 240 000 ($N=20$), 120 000 ($N=22$),
48 000 ($N=24$), 10 000 ($N=26$), 3 000 ($N=28$), 914 ($N=30$), 516 ($N=32$), 90 ($N=34$).
}
  \label{fig:g-majorana-beta0}
\end{figure}

Figure~\ref{fig:g-majorana-beta0} shows $g(t)$ at $\beta=0, 1, 5$ for various values of $N$.
The corresponding RMT ensembles are
\begin{center}
\begin{tabular}{c|c|c|c|c|c|c|c|c|c|c}
$N$ & 16 & 18 & 20 & 22 & 24 & 26 & 28 & 30 & 32 & 34\\
\hline
class & GOE & GUE & GSE & GUE & GOE & GUE & GSE & GUE & GOE & GUE
\end{tabular}
\end{center}
The shape of the ramp in each case agrees with the RMT prediction outlined above.
In particular, the kinks visible for $N=20,28$ are a signature feature of the ramp in the GSE ensemble.   
For $N=34$ (GUE) a careful comparison that confirms the RMT ramp shape is described in Section \ref{rampSYK}. As an initial test we fitted the ramp at times well before the plateau time (where unfolding effects discussed in Section \ref{rampSYK} become significant).  We found a power behavior agreeing with the expected GUE behavior $g(t) \sim t^{1}$ to within a few percent.  
These are strong pieces of evidence  that the ramp structure in SYK can be attributed to random matrix theory.

For $\beta=0$ the early time behavior exhibits oscillations, which will not play a role in this work.
The oscillations are due to the fact that, at infinite temperature, the spectral form factor is sensitive to the hard edges at both ends of the energy spectrum.

Let us now consider the plateau heights of Figure~\ref{fig:g-majorana-beta0} in detail.
They are equal to the time-average value of $g(t)$, which at $\beta=0$ is given by \eqref{timeavgDeg}
\begin{align}
  \frac{\sum_E N_E^2}{L^2} \ed
\end{align}
Here $N_E$ is the degeneracy of energy level $E$.
As explained in Appendix~\ref{P}, the SYK spectrum has a double degeneracy ($N_E=2$) when $(\NmodEight) \ne 0$ due to the particle-hole symmetry, leading to a plateau height of $2/L$ at $\beta=0$.
When $(\NmodEight) = 0$ there is no protected degeneracy, and in those cases the plateau height is $1/L$.
These facts are consistent with the pattern of plateau heights exhibited by Figure~\ref{fig:g-majorana-beta0}.
In particular, they explain why the plateaus of $N=16,24,32$ are reduced by a factor of 2 compared with the rest.

One important consequence of Figure~\ref{fig:g-majorana-beta0} is that it allows us to learn about the large $N$ behavior of the ramp.
As we go to larger $N$ the dip time grows quickly, but the plateau time grows even faster, resulting in a more and more prominent ramp.
(For further discussion of the numerical evidence, see Appendix~\ref{App:dtime}.)
We are led to the reasonable conjecture that the ramp is a feature of the large $N$ theory, and that the dip time is a new time scale in the theory.
In Section~\ref{rampSYK} we will present an analytic argument that supports this conclusion.

\section{Thermodynamics of the SYK model}
\label{thermo}

In this section we compute the thermodynamic properties of SYK numerically, and extrapolate to the large $N$ limit.
We find excellent agreement with existing analytic results, both for the infinite $N$ limit and for the leading $O(1/N)$ correction.
This serves as an important cross-check both on our results and on existing results.

We begin with a brief review of the known analytic results.
There is an exact rewrite of the SYK model in terms of bi-local anti-symmetric variables $G(\tau_1,\tau_2)$ and $\Sigma(\tau_1,\tau_2)$.
The path integral is given in Euclidean time by \cite{KitaevTalks,Sachdev:2015efa}
\begin{align}
\label{gsigint}
  Z &= \int \! DG D\Sigma \, e^{-I} \ec \\
  \frac{I}{N} &= - \frac{1}{2} \log \det (\dho_\tau - \Sigma) + \frac{1}{2}
  \int_0^\beta d\tau_1 d\tau_2 \left[ 
  \Sigma(\tau_1,\tau_2) G(\tau_1,\tau_2) - \frac{J^2}{q} G^q(\tau_1,\tau_2)
  \right] \ed \label{SGS}
\end{align}
We remind the reader that we set $q=4$ in most of our analysis.
The action \eqref{SGS} is obtained by performing the disorder average over couplings $J_{ijkl}$, introducing a Hubbard-Stratonovich field for the fermion bi-linear, and integrating out the fermions \cite{KitaevTalks,Sachdev:2015efa}.
In particular, $G(\tau_1,\tau_2)$ should be thought of as the fermion bi-linear $\frac{1}{N} \sum_{a=1}^N \psi_a(\tau_1) \psi_a(\tau_2)$  and $\Sigma(\tau_1, \tau_2)$ as a Lagrange multiplier enforcing this identification.   To compute $\langle Z(\beta +it)Z(\beta -it)\rangle_J$ we need two copies (called replicas) of the fermion fields labelled by replica indices $\alpha, \beta = 1, 2$.  $G, \Sigma$ become $G_{\alpha\beta}, \Sigma_{\alpha\beta}$. The convergence of (\ref{gsigint}) is manifest with a contour choice described in Appendix \ref{toyapp}.

To solve the theory at large $N$ one now writes the saddle point equations for the bi-local fields.
\begin{align}
  \frac{1}{G(\omega)} &= -i\omega - \Sigma(\omega) \ecq
  \Sigma(\tau) = J^2 G^{q-1}(\tau) \ed \label{spEq}
\end{align}
The first equation is in frequency space, and the second is in Euclidean time.
These equations can be solved analytically in the limits $\beta J \to 0$ and $\beta J \to \infty$ \cite{parcollet1999non}, and can be solved numerically for arbitrary values of $\beta J$.
Plugging the result back in \eqref{SGS} gives the large $N$ thermal free energy \cite{PhysRevB.63.134406}.
Certain perturbative $1/N$ corrections to the free energy have also been computed \cite{PolchStreich,Maldacena:2016hyu}.

At finite $N$ we compute the mean energy and other thermodynamic quantities numerically by fully diagonalizing the Hamiltonian.
To make contact with the analytic calculation, we extrapolate the numerical results to large $N$ as follows.
At fixed temperature $T$ we compute $\langle E(T) \rangle/N$ at different $N$ values and fit to a polynomial in $1/N$ of degree 2.
The leading $O(N^0)$ coefficient is then the infinite $N$ result, the next term is the $1/N$ correction, and so on.

\begin{figure}[h]
  \centering
  \includegraphics[width=0.7\textwidth]{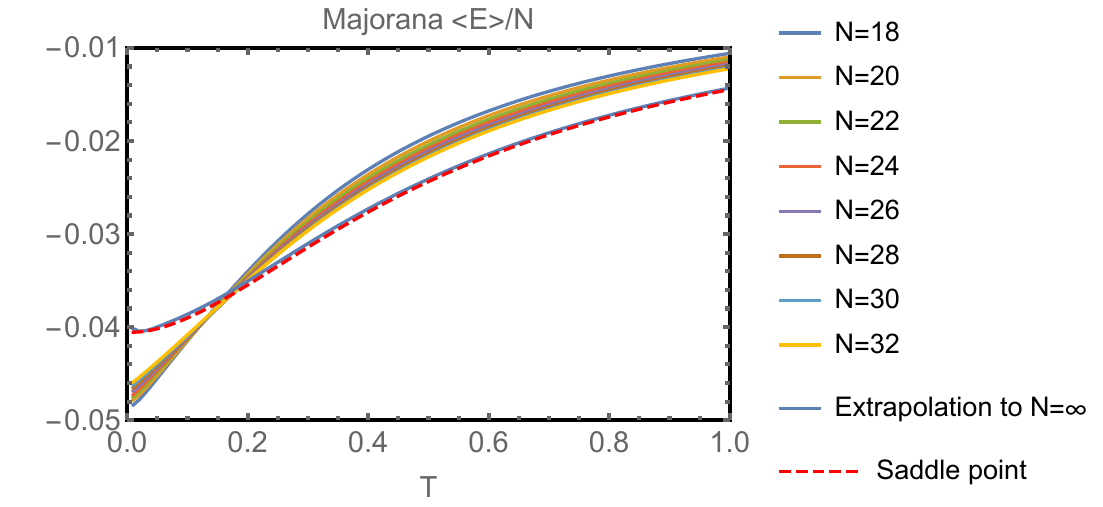}
  \caption{Shown are SYK thermodynamic $\langle E(T) \rangle /N$ for different values of $N$, computed by exact diagonalization. We also plot the point-wise extrapolation obtained by fitting the eight values of $N$ to a three-tem expansion in $1/N$ and taking the leading term. This is almost indistinguishable from the exact large $N$ result obtained by solving the Schwinger-Dyson equations numerically.
}
  \label{fig:majorana-energy-vs-T.pdf}
\end{figure}
Figure~\ref{fig:majorana-energy-vs-T.pdf} shows the mean energy extrapolated to infinite $N$, compared with the result obtained from a direct solution of the large $N$ saddle point equations.
We find excellent agreement between the two methods of computation, although even at $N=32$ (the largest value considered here) the result is not close to the infinite $N$ answer.
The mean energy can be written at low temperature as
\begin{align}\label{Etherm}
  \langle E(T) \rangle = N \epsilon_0 + a T + \frac{N c}{2} T^2 + N c_2 T^3+\cdots
\end{align}
The normalization is such that all coefficients scale as $O(N^0)$.
The coefficients $c$ (the large $N$ specific heat) and $a$ have been computed in the large $N$ theory\footnote{
The coefficient $a$ was computed in \cite{PolchStreich,Maldacena:2016hyu} from a one-loop fluctuation correction to the large $N$ saddle, or equivalently from summing diagrams formed by bending ladder diagrams around into a loop.
}
and are given by
\begin{align}
\epsilon_0 \approx -0.0406 \ecq
  a = \frac{3}{2} \ecq
  \frac{c}{2} \approx 0.198 \ed \label{acoeffs}
\end{align}
The coefficient $c_2$ has not been reported in the literature, but we believe it should be $c_2=-0.419$.\footnote{This is based on a conjectured $1/\beta^2$ term in the free energy, which in the notation of \cite{Maldacena:2016hyu} reads
\be
\frac{\log Z}{N} = \#\beta \mathcal{J} + s_0 + \frac{2\pi^2\alpha_S}{\beta\mathcal{J}} - \frac{2\pi^2\alpha_S\alpha_K}{(\beta\mathcal{J})^2|k'_c(2)|}+...
\ee}
Notice that the linear term in \eqref{Etherm} is subleading in $1/N$.
This must be the case because this term corresponds to a $\log(T)$ term in the entropy, which becomes negative at finite temperature.
Let us now compare these coefficients to the extrapolated numerical results:\footnote{We included a $T^{3.77}$ term to account for the first nontrivial operator dimension in the model \cite{Maldacena:2016hyu}. Surprisingly, the fit agrees with large $N$ results slightly better if we replace this with a $T^4$ term.}
\begin{align}
\frac{E}{N} =   -0.04 - 0.0025 T + 0.22 T^2 - 0.52 T^3 + 0.37 T^{3.77} \ed
\end{align}
 We see that $a$ is suppressed at infinite $N$ as expected, while $c$ is within fifteen percent of the expected value \eqref{acoeffs}.
Next, we fit the $1/N$ piece of the extrapolated energy and find
\begin{align}
  -0.23 + 1.6 T - 3.4 T^2 + 2.9 T^3 \ed
\end{align}
Here the fitted value of $a = 1.6$ is fairly close to the expected value $a = \frac{3}{2}$.

\begin{figure}[h]
  \centering
  \includegraphics[width=0.7\textwidth]{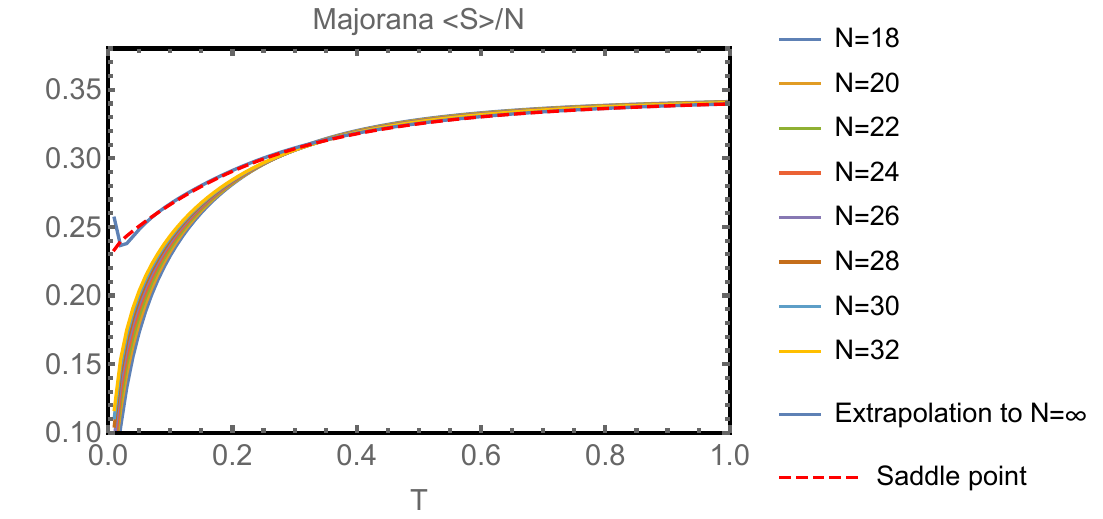}
  \caption{SYK thermodynamic $S(T)/N$, analyzed in the same way as Fig.~\ref{fig:majorana-energy-vs-T.pdf}.}
  \label{fig:majorana-entropy-vs-T.pdf}
\end{figure}

Next, Figure~\ref{fig:majorana-entropy-vs-T.pdf} shows the entropy extrapolated to infinite $N$.
We again find excellent agreement with a direct infinite $N$ calculation.
At low temperature the entropy is given by
\begin{align}\label{Stherm}
  S(T) &= N s_0 + a \log(T) + N c T + \cdots \ed
\end{align}
Here $s_0 \approx 0.2324 \approx \frac{1}{2} \log(1.592)$ is the analytic zero-temperature entropy density (in the large $N$ limit).
Notice that the numerical extrapolation correctly captures the large $N$ zero-temperature entropy, even though at any fixed $N$ the entropy goes to zero as $T \to 0$.

\section{Spectral form factor in random matrix theory}
\label{rampRMT}

In this section we review properties of the spectral form factor in the GUE random matrix ensemble \cite{mehta2004random,Guhr:1997ve}.
We derive two of the main features of Figure~\ref{fig:g-RMT}: the late time behavior of the slope and the early time behavior of the ramp.
Both are described by power laws, and from there we get an estimate of the dip time in RMT.

Consider the GUE ensemble of Hermitian matrices $M$ of rank $L$, with ensemble averaging defined by
\begin{align}
  \cZ_{\GUE} = \int \prod_{i,j} d M_{ij} \exp{\left( -\frac{L}{2} \trace (M^2) \right)} \ed \label{ZGUE}
\end{align}
In this context, the matrix $M$ is analogous to the SYK Hamiltonian, and the rank $L$ corresponds to the dimension of the Hilbert space.
One important difference is that the natural perturbative parameter in SYK is $1/N$, whereas in RMT we typically expand in $1/L \sim e^{-N}$.

The partition function for a given realization of $M$ is defined by
\begin{align}
  Z(\beta,t) \equiv \trace \left( e^{-\beta M - iM t} \right) \ed \label{RMT}
\end{align}
The spectral form factor $g$ and the related quantities $g_d$ and $g_c$ are then defined by \eqref{g}-\eqref{gc}, where the average $\langle \cdot \rangle_J$ over the random couplings is replaced by the average $\langle \cdot \rangle_{\GUE}$ over random matrix elements, given by \eqref{ZGUE}.\footnote{
For example, for the partition function we have
\begin{align}
  \langle Z(\beta,t) \rangle_{\rm GUE} = 
  \frac{1}{\cZ_{\GUE}}
  \int d M_{ij} \, e^{-\frac{L}{2} \trace (M^2)}
  \trace \left( e^{-\beta M - iM t} \right) \ed
\end{align}
}

Let us diagonalize $M$ and change variables from its matrix elements to its eigenvalues and a unitary change of basis.
This introduces a Jacobian that describes repulsion between eigenvalues.
In the large $L$ limit the eigenvalues can be described by a density $\rho$. We will use $\rho$ for the physical density, and $\tilde{\rho}$ for the unit normalized density:
\be
\int d\lambda \rho(\lambda) = L, \hspace{20pt} \int d\lambda \tilde{\rho}(\lambda) = 1, \hspace{20pt} \rho(\lambda) = L\tilde{\rho}(\lambda).
\ee
Replacing the individual eigenvalues $\lambda_i$ by $\tilde{\rho}(\lambda)$, one obtains\footnote{
  The normalization of $\rho(\lambda)$ should be imposed (for example) by a Lagrange multiplier.
  The resulting saddle point equations are solved subject to this constraint.
  }
\begin{align}
  \cZ_{\GUE} = \int \! D\tilde{\rho}(\lambda) \, e^{-S} \ecq
  S = 
  - \frac{L^2}{2} \int \! d\lambda \, \tilde{\rho}(\lambda) \lambda^2
  + L^2 \int \! d\lambda_1 d\lambda_2 \, 
  \tilde{\rho}(\lambda_1) \tilde{\rho}(\lambda_2) \log |\lambda_1-\lambda_2|
  \ed \label{ZGUEs}
\end{align}
The large $L$ saddle point of the above is given by the Wigner semicircle law,
\begin{align}
  \langle \tilde{\rho}(\lambda) \rangle_{\GUE} =
  \tilde{\rho}_s(\lambda) \equiv \frac{1}{2\pi} \sqrt{4-\lambda^2} \ed
  \label{semicircle}
\end{align}
The physical eigenvalue density is given by $\langle \rho(\lambda)\rangle = L\tilde{\rho}_s(\lambda)$. Notice that the average eigenvalue spacing is of order $1/L$.  

We now turn to discuss the slope and ramp that appear in the spectral form factor, shown in Figure~\ref{fig:g-RMT}.
Roughly speaking, $g(t)$ is dominated by the disconnected piece $g_d(t)$ before the dip time, and by the connected piece $g_c(t)$ after the dip time.
We will discuss each in turn.

The leading large $L$ behavior of $Z(\beta,t)$ follows from the semicircle law.
Working for simplicity at infinite temperature, we have
\begin{align}
  \langle Z(\beta=0,t) \rangle_{\GUE} =
  \int_{-2}^2 d\lambda L \tilde{\rho}_s(\lambda) e^{-i\lambda t} =
  \frac{L J_1(2t)}{t} \ed
\end{align}
Here $J_1$ is a Bessel function of the first kind.
At late times we find that the partition function decays as $L /t^{3/2}$,
and therefore at late times we have
\begin{align}
  g_d(t) &\equiv \frac{\left| \langle Z(0,t) \rangle_J \right|^2}{L^2} 
  \sim \frac{1}{t^3} \ed
  \label{gdRMTslope}
\end{align}
This is true also at finite temperature.
Before the dip time, the spectral form factor $g(t)$ is dominated by the disconnected part $g_d(t)$.
Therefore, the late time decay of $g(t)$ before the dip time is also proportional to $1/t^{3}$. This particular power is a consequence of the fact that the mean eigenvalue density (\ref{semicircle}) vanishes as a square root near the edge of the spectrum.

\subsection{The ramp and the dip time}

We now review how to derive the presence of a ramp in RMT. We focus for simplicity on the connected spectral form factor $g_c(t;\beta=0)$, and show that $g_c(0,t) \sim \frac{t}{L^2}$ at times $1 \ll t \le L$.
Beyond the dip time, $g(t)$ and $g_c(t)$ are almost equal, both exhibiting the ramp/plateau structure.
However, for $g_c$ the ramp extends to very early times, giving better perturbative control.\footnote{In fact, $1/L$ perturbation theory remains valid up to times $t \sim \epsilon L$ where $\epsilon$ is a small $L$-independent parameter.}

The connected spectral form factor can be written as
\begin{align}
  g_c(t;0) &= \int \! d\lambda_1 d\lambda_2 \, R_2(\lambda_1, \lambda_2) 
  e^{i(\lambda_1 - \lambda_2) t} \ec \label{R21}\\
  R_2(\lambda_1,\lambda_2) &\equiv 
  \langle \delta \tilde{\rho}(\lambda_1) \delta \tilde{\rho}(\lambda_2) \rangle_{\GUE}
  \ed \label{R2}
\end{align}
Here $R_2$ is the connected pair correlation function of the unit-normalized density $\tilde{\rho}$, and $\delta \tilde{\rho}(\lambda) \equiv \tilde{\rho}(\lambda) - \tilde{\rho}_s(\lambda)$ is the fluctuation around the mean eigenvalue density $\tilde{\rho}_s(\lambda)$ given by the semicircle law \eqref{semicircle}.
A basic result of RMT is that, near the center of the semicircle, 
$R_2(\lambda_1,\lambda_2)$ is given by the square of the \textit{sine kernel} \cite{mehta1960statistical,gaudin1961loi,dyson1962statistical} plus a delta function at coincident points:
\begin{align}
  R_2(\lambda_1,\lambda_2) = - \frac{
  \sin^2 \left[ L (\lambda_1 - \lambda_2) \right]}{
  \left[ \pi L (\lambda_1 -\lambda_2)\right]^2}  + \frac{1}{L\pi}\delta(\lambda_1-\lambda_2). \label{sinr}
\end{align}
Fourier transforming as in (\ref{R21}) gives
\begin{align}
  g_c(t) \sim 
  \left\{ \begin{matrix}
    t/(2\pi L^2) \ec & \, t < 2 L \\
    1/(\pi L) \ec & \, t \ge 2 L
  \end{matrix} \right. \ed \label{gcRMT}
\end{align}
This explains the observed behavior in Figure~\ref{fig:g-RMT} beyond the dip time: There is a ramp up to the plateau time $2 L$, and a constant plateau value beyond.\footnote{
  Our analysis here only applies to the contribution from eigenvalues near the center of the semicircle, where the mean density is $L/\pi$. We will show how to include regions with different mean densities in (\ref{ZZbargeneral}).  Brezin and Hikami derive  remarkable nonperturbative formulas for $g(t)$ in \cite{brezinhikami,Brezin:1997zz}. 
  } The ramp lies below the plateau because the eigenvalues are anticorrelated as reflected in the minus sign in \eqref{sinr}.

This is a good explanation of the ramp and plateau, but it requires an appeal to the sine kernel. In fact, one can derive the ramp in a more basic way without knowing about the sine kernel. Notice that the initial linear time dependence of the ramp can be obtained by approximating the sine kernel by
\begin{align}
  R_2(\lambda_1,\lambda_2) \approx - \frac{1}{2(\pi L(\lambda_1 - \lambda_2))^2} \ed \label{R2approx}
\end{align}
We now review how to derive this perturbatively from the action (\ref{ZGUEs}) following Altshuler and Shklovskii \cite{altshuler1986repulsion}. Writing $\tilde{\rho} = \tilde{\rho}_s + \delta \tilde{\rho}$ and expanding the action \eqref{ZGUEs} about the saddle point, we find the quadratic term
\begin{align}
  \delta S = - L^2 \int \! d\lambda_1 \, d\lambda_2 \,
  \delta \tilde{\rho}(\lambda_1) \delta \tilde{\rho}(\lambda_2) \log|\lambda_1 - \lambda_2| \ed
\end{align}
We can now carry out the Gaussian integral to determine the two-point function \eqref{R2}.
We go to Fourier space $\delta \tilde{\rho}(\lambda) = \int \! \frac{ds}{2\pi} \, \delta \tilde{\rho}(s) \exp(i s \lambda)$ and find
\begin{equation}
  \delta S = \frac{L^2}{2} \int \! ds \, 
  \delta \tilde{\rho}(s) \frac{1}{|s|} \delta \tilde{\rho}(-s) \ed
\end{equation}
Notice that long-wavelength fluctuations of $\rho$ are strongly suppressed:  This is the spectral rigidity referred to in RMT.\footnote{By observing that the local eigenvalue density is the inverse of the level spacing one can read off from the following result the $\langle \delta E_n \delta E_m \rangle \sim \log |E_n -E_m|$   signature of spectral rigidity discussed earlier.}
Then we find 
\begin{align}
  \langle \delta \tilde{\rho}(\lambda_1) \delta \tilde{\rho} (\lambda_2) \rangle 
  &= 
  \frac{1}{4\pi^2L^2}
  \int ds e^{i(\lambda_1 - \lambda_2) s} |s| + O(L^{-4})
  \cr
  &= - \frac{1}{2(\pi L(\lambda_1 -\lambda_2))^2} + O(L^{-4})\ed\label{pertResult}
\end{align}

A calculationally more efficient way of studying $g(t)$ in RMT is the formalism developed by Brezin and Zee \cite{Brezin:1993qg} which uses standard `t Hooft large $L$ perturbation theory to compute the double resolvent of $M$.   We discuss this technology in Appendix \ref{Nq}.

Equating the slope \eqref{gdRMTslope} and the ramp \eqref{gcRMT} gives the RMT dip time $t_d \sim \sqrt{L}$, exponential in the ``entropy" $\log L$. We find that the ratio $t_p / t_d \sim \sqrt{L}$, also exponential in the entropy, and therefore the ramp in the RMT spectral form factor survives in the large $L$ limit.  

This derivation makes it clear that (\ref{pertResult}) is a perturbative result in RMT at order $1/L^2$. Its contribution to $g_c(t)$ is proportional to $t/L^2$, capturing the ramp part of \eqref{gcRMT}. In other words the ramp is a perturbative RMT effect. By contrast, the plateau is not.\footnote{
  In GUE the ramp is the full perturbative result, while in other RMT ensembles (such as GOE and GSE) the ramp receives higher-order perturbative corrections.
  Non-perturbative corrections to the ramp exist in all cases.
  }
  Indeed, the appearance of the plateau depends on the oscillating factor in the more exact sine kernel \eqref{sinr} expression, which can be obtained from a RMT instanton expression $e^{-2LE_{\rm imag}}$ with imaginary energy \cite{andreev1995spectral, kamenevmezard}. 
    The oscillating term comes from continuing to real energy and extracting the appropriate part of the result.

This has important consequences for the application to the SYK model. For SYK, $L = 2^{N/2}$ so $1/L \sim e^{-aN}$.
  Therefore, perturbative RMT effects are nonperturbative in SYK, of order $e^{-2aN}$.   Nonperturbative RMT effects of order $e^{-L}$ are of order $\exp(-e^{aN})$, an extremely small nonperturbative effect.

\section{Spectral form factor in the SYK model}
\label{rampSYK}

The presence of the ramp in the results of Section~\ref{spectral} suggests that the SYK model possesses spectal rigidity, even for eigenvalue spacings far larger than the mean nearest-neighbor spacing. By combining this assumption with coarse-grained features of the large $N$ spectrum, we reproduce reasonably well the $g(t)$ curve obtained from exact diagonalization.

First, let us explain how an assumption of spectral rigidity produces the ramp observed in $g(t)$. Starting with the general definition of $\langle Z Z^*\rangle$,

\be
\langle Z(\beta+it)Z(\beta-it)\rangle = \int d\lambda_1 d\lambda_2 \langle \rho(\lambda_1)\rho(\lambda_2)\rangle e^{-\beta(\lambda_1+\lambda_2)}e^{-i(\lambda_1-\lambda_2)t},
\ee
it is convenient to define $x = \lambda_1-\lambda_2$ and $E = \frac{\lambda_1+\lambda_2}{2}$. Notice that in this expression and below, we are using $\rho$, the physical eigenvalue density, normalized so $\int \! d\lambda \, \rho = L$. 

Now, for late times we assume that the integral is dominated by regions where $x$ is sufficiently small that we can approximate the density-density correlator by GOE, GSE, or GUE statistics. For simplicity, we take GUE statistics
\be
\langle \rho(\lambda_1)\rho(\lambda_2)\rangle = \langle\rho(E)\rangle \delta(x) + \langle \rho(\lambda_1)\rangle\langle \rho(\lambda_2)\rangle\left(1 - \frac{\sin^2\left[\pi\langle \rho(E)\rangle x\right]}{\left[\pi\langle \rho(E)\rangle x\right]^2}\right),
\ee
which leads to\footnote{We should make a few comments about this formula. First, if the local statistics are GOE or GSE, then we would replace the ramp function in (\ref{ZZbargeneral}) by the appropriate spectral form factor. Second, in cases where the spectum is uniformly $d$-fold degenerate, we should multiply the ramp term by $d^2$ and divide $\langle \rho(E)\rangle$ inside the second term by $d$.}
\be\label{ZZbargeneral}
\langle Z(\beta+it)Z(\beta-it)\rangle = |\langle Z(\beta+it)\rangle|^2 + \int dE\, e^{-2\beta E} \text{min}\left\{\frac{t}{2\pi},\langle \rho(E)\rangle\right\}.
\ee

Eq.~(\ref{ZZbargeneral}) can be interpreted as follows: we approximate the spectrum by bands over which $\rho(E)$ varies very little. From each band, we get a GUE ramp. The integral over energy in (\ref{ZZbargeneral}) is simply summing up these individual ramps which then yields another smoothed ramp. This is the inverse of an ``unfolding'' process.  In a theory with many degrees of freedom, we expect the integral over $E$ to be strongly peaked around a maximum. In general, the location of this maximum will depend on time. The ramp will join the plateau at the time $t_p = e^{S(2\beta)}$, where the energy that maximizes the integral is simply $E(2\beta)$, the energy that dominates the canonical ensemble at inverse temperature $2\beta$. One can check that the derivative of $g_c(t)$ will smoothly approach zero at $t_p$, giving a $C^1$ transition onto the plateau even though individual bands have a kink.

One would like to apply (\ref{ZZbargeneral}) to SYK, but there is an important subtlety. The second term in (\ref{ZZbargeneral}) should be understood as exponentially smaller than the first, as long as $t$ is not too large. In SYK, we also expect correlations between eigenvalues that are only power-law suppressed by $N$ (more precisely of order $1/N^q$). One source of such fluctuations would be the overall scale of the Hamiltonian, which varies from $J$ configuration to $J$ configuration. Such terms would dominate over the ramp contribution at short times. However, we might hope that these $1/N^q$ terms will always be smaller than either the first term or the second term in (\ref{ZZbargeneral}), so the formula still gives a reasonable picture of SYK. We will return to this point below.

Let us now attempt to evaluate (\ref{ZZbargeneral}) for large $N$ SYK. First we discuss the disconnected first term. We can numerically evaluate the large $N$ saddle point that determines $\langle Z(\beta+it)\rangle$, but for large values of $\beta +it$, we also need to consider fluctuations about this saddle. There are a set of modes that become soft for large $\beta+it$, which can be captured by the partition function of the effective Schwarzian derivative theory \cite{kitaevfundamental,Maldacena:2016hyu}:
\begin{equation}\label{schaction}
Z_{Sch}(\beta) = \int\frac{D[\tau(u)]}{SL(2,R)}\exp\left[-\frac{\pi N\alpha_S}{\beta\mathcal{J}}\int_0^{2\pi} du\left(\frac{\tau''^2}{\tau'^2} - \tau'^2\right)\right].
\end{equation}
Here, $0<u<2\pi$ is the physical time variable of the model, and $\tau(u)$ is a reparametrization of the thermal circle. The parameter $\mathcal{J}$ sets the scale of the Hamiltonian in a way appropriate for general values of $q$, and $\alpha_S$ is a numerical coefficient that depends on $q$; these are related to the specific heat $c$ by $c = \frac{4\pi^2\alpha_S}{\mathcal{J}}$. The classical and one-loop contributions to this action have been studied previously \cite{Maldacena:2016hyu}, with the result
\be\label{oneloopresult}
Z_{Sch}^{1-loop}(\beta) = \frac{\#}{(\beta\mathcal{J})^{3/2}}\exp\left(\frac{2\pi^2 N\alpha_S}{\beta\mathcal{J}}\right).
\ee
However, notice that when we continue to large values of $\beta+it$, the coefficient multiplying the action (\ref{schaction}) becomes small, and $\tau(u)$ will have large fluctuations. Naively, this invalidates a perturbative analysis, making it difficult to evaluate $Z$. In fact, with the correct measure, the theory turns out to be one-loop exact. We will present a somewhat indirect derivation of this fact. A direct proof is also possible \cite{Stanford:2017thb}.

Our derivation is based on an intermediate step where we think about the SYK model for large values of $q$. Then the coefficient in the action becomes \cite{Maldacena:2016hyu}
\be
\frac{\pi N\alpha_S}{\beta\mathcal{J}}\rightarrow \frac{\pi}{4\beta\mathcal{J}}\cdot\frac{N}{q^2}.
\ee
In particular, the coefficient is only a function of $\frac{N}{q^2}$. We can therefore take a ``double-scaled'' limit of large $N$ and large $q$ with $\frac{N}{q^2}$ held fixed. It is clear that the Schwarzian part of the theory will survive in this limit, but the rest of the SYK theory simplifies significantly, and it becomes possible to exactly compute the disorder-averaged density of states using techniques from \cite{erdHos2014phase}. We sketch this in Appendix~\ref{doubleScaledSYK}.\footnote{
  Notice that if we take a double scaling limit $q,N \to \infty$ keeping $q/N^\alpha$ fixed, then the scrambling time is of order $\log(N)$ when $\alpha < \frac{1}{2}$, while it is of order 1 when $\alpha > \frac{1}{2}$.
  Therefore $q^2 \sim N$ marks the boundary between the behaviors expected for k-local and nonlocal Hamiltonians \cite{Sekino:2008he, Hayden:2007cs, Lashkari:2011yi}.
}
To isolate the contribution of the Schwarzian, we take a further ``triple-scaled'' limit where we take $\frac{N}{q^2}$ large and the energy $(E-E_0)$ above the ground state small, with the product held fixed. In this limit, we find the density of states (see Appendix~\ref{doubleScaledSYK} eq.~(\ref{sinhtriple}))
\be\label{sinhdensity}
\rho(E) \propto \sinh\left(\pi\sqrt{2z}\right), \hspace{20pt} z = \frac{(E-E_0)N}{q^2\mathcal{J}}\rightarrow\frac{4\alpha_S(E-E_0) N}{\mathcal{J}} = \frac{c(E - E_0)N}{\pi^2}.
\ee
We expect that the Schwarzian sector is the only part of the theory that survives this triple-scaled limit, so (\ref{sinhdensity}) should be an exact result for the Schwarzian theory. Computing the partition function via $Z = \int dE \rho(E) e^{-\beta E}$, we learn that the one-loop result (\ref{oneloopresult}) is actually the exact answer for the Schwarzian theory.

The conclusion of this discussion is that we can include the effect of the soft mode integral by simply dividing the large $N$ saddle point expression for the partition function by a factor of $(\beta + it)^{3/2}$. Using the expression for the large $N$ free energy in the holographic limit, $\log Z = N(\epsilon_0\beta + s_0 + \frac{c}{2\beta})$, one finds that the disconnected term in (\ref{ZZbargeneral}) contributes the following to $g(t)$:
\be\label{slopelargeN}
\frac{|\langle Z(\beta+it)\rangle|^2}{\langle Z(\beta)\rangle^2} = \frac{\beta^3}{(\beta^2 + t^2)^{3/2}}\exp\left(-\frac{cNt^2}{\beta(\beta^2+t^2)}\right).
\ee
The time dependence of the exponent becomes negligible at $t \gtrsim \sqrt{N}$, and we have a power law decay $\sim t^{-3}$. Away from the holographic limit, one would replace the piece in the exponential by the appropriate finite $\beta$ saddle point action, which can be computed numerically.

Now, we would like to evaluate the second term in (\ref{ZZbargeneral}). Away from the holographic limit, one has to use the numerical $S(E)$ determined by solving the Schwinger-Dyson equations. However, we can give a simple formula in the holographic limit, where $S(E) = N s_0 + \sqrt{2c(E-E_0)N}.$ Neglecting one-loop factors from the integral over $E$, we have the contribution to $g(t)$
\begin{equation}\label{ramplargeN}
g_{ramp}(t) \sim \begin{cases} 
      \frac{t}{2\pi}\exp\left[-2 N s_0 - \frac{cN}{\beta}\right] \ec & \frac{t}{2\pi}< e^{N s_0} \\
      \frac{t}{2\pi} \exp\left[-2N s_0 - \frac{cN}{\beta}-\frac{\beta}{cN}\log^2\left(\frac{t/(2\pi)}{e^{N s_0}}\right)\right] \ec & e^{N s_0}< \frac{t}{2\pi}< \frac{t_p}{2\pi} \\
      \exp\left[-N s_0 -  \frac{3cN}{4\beta} \right] \ec & t_p< t. 
   \end{cases}
\end{equation}
where $t_p = 2\pi e^{Ns_0 + \frac{cN}{2\beta}} = 2\pi e^{S(\beta)}.$ Notice that this function is $C^1$. We can evaluate the dip time by equating (\ref{slopelargeN}) and (\ref{ramplargeN}), which gives $t_d \sim e^{N s_0/2}$.

\begin{figure}[h]
  \centering
\includegraphics[width=0.6\textwidth]{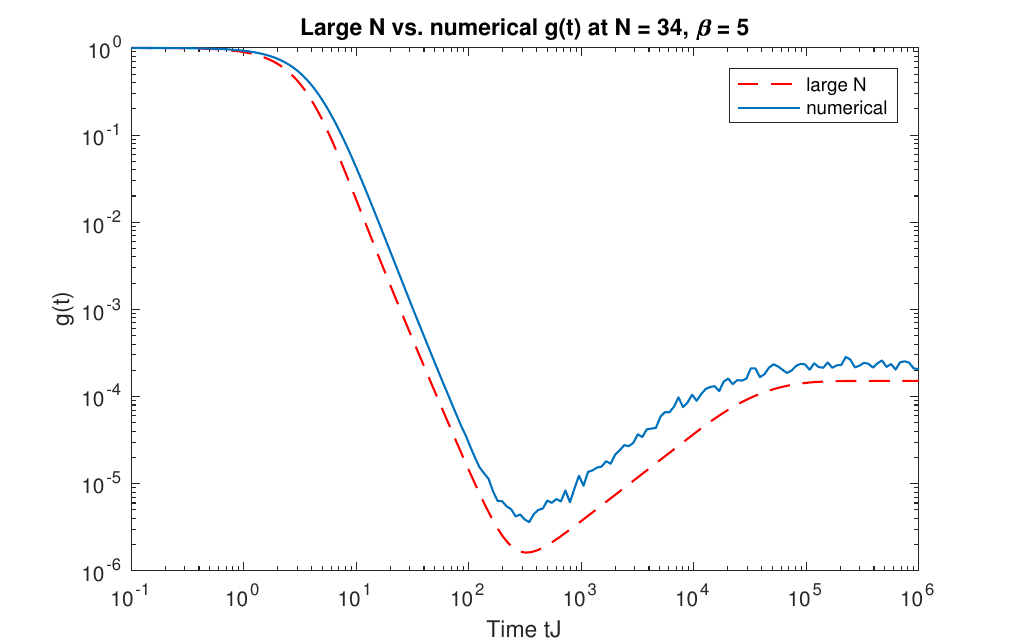}
  \caption{Comparision of (\ref{ZZbargeneral}) evaluated for the SYK model using the large $N$ density of states extrapolated to $N = 34$ plus the Schwarzian partition function (red/dashed) against the $N = 34$ exact diagonalization answer for $g(t)$ (blue/solid). The discrepancies in the ramp and plateau regions are due to the fact that the large $N$ free energy (without a proper one-loop term) gets the partition function wrong by an order one factor. In the ramp we are dividing by this twice, and in the plateau we are roughly dividing by it once.}
  \label{fig:largeNg}
\end{figure}

One can also make a more exact analysis of the large $N$ function, by evaluating the finite $\beta$ saddle point action numerically, and doing the integral over $E$ in (\ref{ZZbargeneral}). In Fig.~\ref{fig:largeNg} we show the result of doing this computation and plugging in $N = 34$ to compare to the exact diagonalization data. We also take into account the two-fold degeneracy in the spectrum for $N = 34$ and evaluate the numerical finite temperature saddle for the slope portion, slightly correcting (\ref{slopelargeN}). The agreement is reasonably good, although the ramp and plateau are off by factors that represent the discrepancy in the exact free energy vs. the large $N$ saddle point. (Presumably this factor would be mostly resolved by a complete one-loop correction to the large $N$ partition function.)

We caution the reader that although the agreement in Fig.~\ref{fig:largeNg} looks reasonable, it is very  possible that the true large $N$ answer for $g(t)$ would differ in important ways. In particular, we are not confident that the slope region continues to be described by the simple Schwarzian effective theory out to very long times of order $e^{Ns_0/2}$. Another possibility is that some effect leads to the slope portion of $g(t)$ decreasing more rapidly at an earlier timescale. For example, this could be the result of some $1/N^q$ effect that tends to smooth out the sharp $\sqrt{E-E_0}$ edge in the spectrum, leading to a faster decay.\footnote{A simple example  of a $1/N^q$ effect is the sample-to-sample variation of the edge of the eigenvalue spectrum.  This causes a gaussian crash in the partition function $g_d(t)$ at times of order $N^{(q-2)/2}$ but cancels out in $g(t)$.   Roughly speaking, effects that cause a crash in $g(t)$ must be present in a single sample.} In this situation, the slope would crash and intersect the ramp much sooner, leading to a short dip time, perhaps of order $t_d \sim N^q$. Another possibility is that the very bottom of the spectrum would be controlled by a spin-glass phase that was argued to exist in the Sachdev-Ye model \cite{PhysRevB.63.134406}, and may also be present at exponentially low temperatures in the SYK model \cite{kitaevSpinGlass}. Such effects may also lead to a softer edge in the spectrum, again leading to a faster decay of the slope and a correspondingly shorter dip time.

We are fairly confident that the dip time should be no {\it later} than $e^{N s_0/2}$, based on the idea that neglected effects are not likely to make the spectrum vanish more sharply. As an extreme fallback position, one can argue without any calculation that the dip time is less than $e^{N s_0}$, which is enough to establish a parametrically long ramp at non-zero temperature. To make the argument, one assumes that the slow decay in the slope is monotonic and roughly independent of temperature, based on the idea that it comes from the edge of the spectrum. Note that $t_d$ can never be larger than $t_p$, because at times $t > t_p$ the spectral form factor $g(t)$ is only sensitive to individual energy levels, with all correlations between different levels getting washed out by the oscillating terms. For $t>t_p$, $g(t)$ is equal to the constant plateau height $g_p$. This allows us to conclude that $t_d(\beta) \le t_d(\beta=\infty) \le t_p(\beta = \infty) = e^{N s_0}$.

\section{Correlation functions}
\label{sec:corr}
In this section we will discuss when two-point correlation functions exhibit ramp + plateau structure at late times.  We will use the  Eigenstate Thermalization Hypothesis (ETH) \cite{deutsch1991quantum,srednicki1994chaos} to estimate  matrix elements.  As we will see, the answer depends on the $(N \text{ mod } 8)$ symmetry pattern \cite{PhysRevB.83.075103,You:2016ldz,Fu:2016yrv}, which is reviewed in Appendix~\ref{P}. 

As before, we focus on the annealed (`factorized') two-point function
\begin{align}
  G(t) \equiv
  \frac{1}{N} \sum_{i=1}^{N} 
  \frac{ 
  \langle \trace \left[ e^{-\beta H} \psi_i(t) \psi_i \right] \rangle_J
  }
  {\langle Z(\beta) \rangle_J}, \label{G}
\end{align}
in which the disorder average is taken separately in the numerator and the denominator.  This quantity is easier to study analytically than the quenched correlator.  We note in passing that it is sometimes useful to consider the average of the squared two-point function \cite{Dyson:2002nt,Barbon:2003aq}, but for our purposes it will be enough to consider the average of the two-point function itself.

Let us first determine whether the two-point function has a nonzero plateau.
This can be determined by considering the following long-time average in a single realization of the random couplings.
\begin{align}
  \frac{1}{Z(\beta)}
  \lim_{t_o \to \infty} \frac{1}{t_o} \int_0^{t_o} dt 
  \trace \left[ e^{-\beta H} \psi(t) \psi(0) \right] =
  \frac{1}{Z(\beta)}
  \sum_{
  \begin{smallmatrix}
    n,m \\ E_n=E_m 
  \end{smallmatrix}
  } 
  e^{-\beta E_n} \left| \langle n | \psi | m \rangle \right|^2 \ed
  \label{Gintt}
\end{align}
Here, $\left| n \right>$ is the energy eigenbasis with energies $E_n$ in the random couplings realization, and $\psi$ stands for any one of the fermions $\psi_i$.
(We neglect the effect of degeneracies for simplicity.)
We expect a non-zero plateau to appear unless the matrix element vanishes.
If $N/2$ is even then there is no degeneracy between the charge parity odd and even sectors (see Appendix~\ref{P}).
In this case the matrix element in \eqref{Gintt} equals zero and the plateau vanishes.

If $N/2$ is odd then we can use the particle-hole operator $P$ to write down a selection rule for the matrix element.
Let $\left| n \right>$, $\left| m \right>$ denote degenerate states with $E_n = E_m$, such that $P \left| m \right> = \left| n \right>$. Then we are interested in whether $\langle m|\psi|n\rangle$ can be nonzero. We have
\be
\langle m|\psi|n\rangle = \Big(|m\rangle,\psi P|m\rangle\Big) = \Big(P\psi P|m\rangle,P|m\rangle\Big) = \eta(N)\Big(\psi|m\rangle,P|m\rangle\Big) = \eta(N)\langle m|\psi|n\rangle,
\ee
where we used inner product notation $\big(|1\rangle,|2\rangle\big) = \langle 1|2\rangle$ for clarity. In the second equality we used the antiunitarity of $P$, and in the third equality we used (\ref{actionOnFermions}).
We conclude that a plateau can only appear when $\eta(N) = 1$, or equivalently when $(N~\mathrm{mod}~8) = 2$.

Next we ask when will a ramp appear in the two-point function.
To answer this question, we put the disorder-averaged two-point function in the following form.
\begin{align}
  G(t) = 
  \frac{1}{N \langle Z(\beta) \rangle_J} \sum_{i}
  \int dE dE' e^{-\beta E} e^{i(E - E')t} \left\langle
  \rho(E) \rho(E') \left| \langle E | \psi_i | E' \rangle \right|^2
  \right\rangle_J \ed \label{Grho}
\end{align}
Here, $\rho(E)$ is the energy spectrum in a given realization of the random couplings.
Notice again that the matrix element $\langle E | \psi_i | E' \rangle$ connects eigenstates from two different charge parity sectors. 
We will again consider two separate cases, depending on whether $N/2$ is even or odd.

If $N/2$ is even then there is no degeneracy between the two sectors.
The two $\rho$ factors that appear in \eqref{Grho} are de-correlated for sufficiently small energy differences (corresponding to late times), and we do not expect a ramp to appear.

If $N/2$ is odd then the two charge parity sectors are degenerate, so effectively there is only one sector.
As discussed above, at late times the correlator probes small energy differences in the spectrum, where we expect each sector of the Hamiltonian to resemble a Gaussian random matrix. 
For such a matrix, the averages over eigenvalues and eigenstates factorize, and we can approximate
\begin{align}
  \left\langle
  \rho(E) \rho(E') \left| \langle E | \psi_i | E' \rangle \right|^2
  \right\rangle_J \approx
  \left\langle \rho(E) \rho(E') \right\rangle_J \cdot
  \left<
  \left| \langle E | \psi_i | E' \rangle \right|^2 \right>_J \ed
\end{align}
Furthermore, for a Gaussian random matrix $\left\langle \left| \langle E | \psi_i | E' \rangle \right|^2 \right\rangle_J$ is a smooth function of the small energy difference $|E-E'|$, as in ETH,  and we approximate it by a constant.
The value of this function at $E=E'$ determines whether there is a non-zero plateau, as discussed above.
The remaining factor $\left\langle \rho(E) \rho(E') \right\rangle_J$ gives the spectral form factor.
It will lead to a ramp, just as in the case of the observable $g(t)$ discussed in previous sections.

\begin{figure}[h]
  \centering
\includegraphics[width=0.4\textwidth]{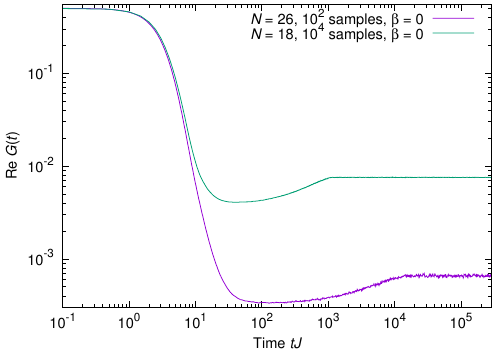}
\includegraphics[width=0.4\textwidth]{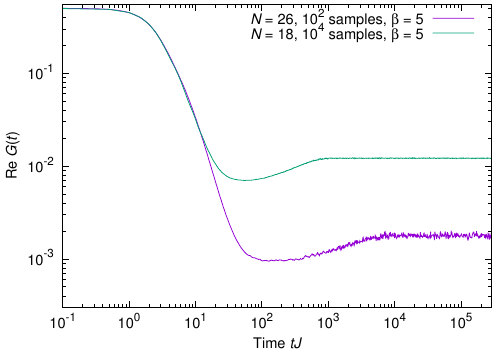}
  \caption{SYK two-point function \eqref{G} for $N = 18, 26$, plotted for $\beta=0,5$.
A slope, dip, ramp, and plateau can be seen.}
  \label{fig:maj2ptN18}
\end{figure}

\begin{figure}[h]
  \centering
\includegraphics[width=0.8\textwidth]{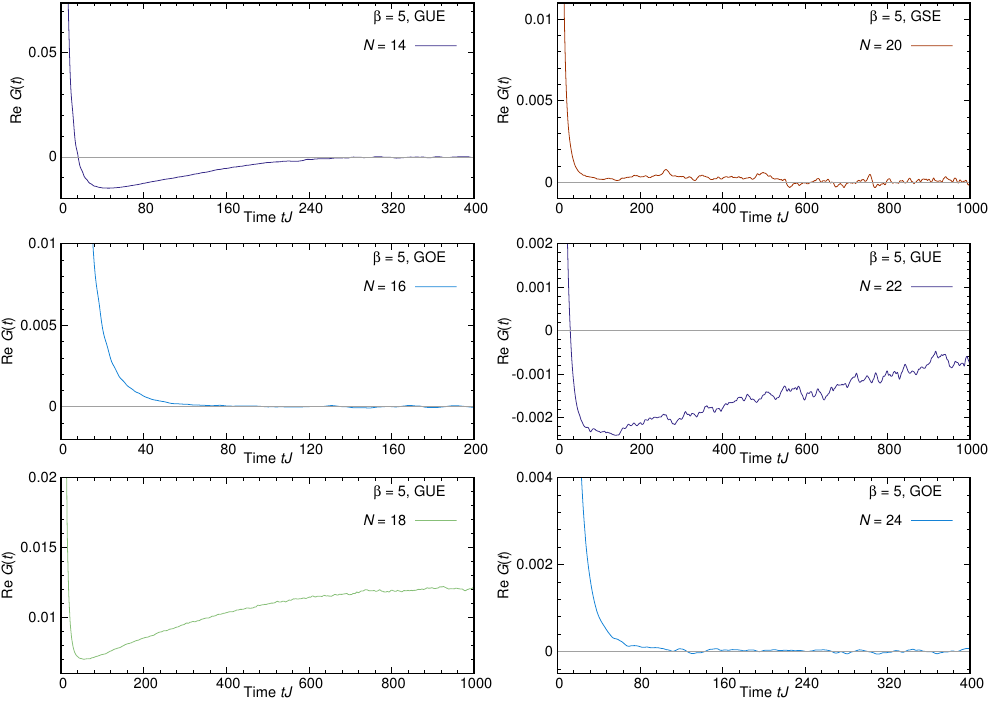}
  \caption{SYK two-point function $G(t)$ for $N = 14, 16, 18$ (left) and
$N = 20, 22, 24$ (right), $\beta=5$.
The number of samples is $10^4$ for $14\leq N\leq 18$, $10^3$ for $N=20$, and $10^2$ for $N=22, 24$.
A ramp appears for $N~\mathrm{mod}~8= 2, 6$ but not for $N~\mathrm{mod}~8= 0, 4$.
    A non-zero plateau appears only for $N~\mathrm{mod}~8 = 2$.
    These properties are all explained by the $(N~\mathrm{mod}~8)$ symmetry pattern.
  } \label{fig:maj2pt}
\end{figure}

To summarize, the two-point function will display the following combinations of a ramp and a non-zero plateau, depending on the value of $(N~\mathrm{mod}~8)$.
\begin{itemize}
  \item If $(N~\mathrm{mod}~8) = 0,4$ then there will be no ramp or plateau.
  \item If $(N~\mathrm{mod}~8) = 2$ then there will be a ramp and a non-zero plateau.
  \item If $(N~\mathrm{mod}~8) = 6$ then there will be a ramp but no plateau (the two-point function will vanish at late times).
\end{itemize}
Figures \ref{fig:maj2ptN18}, \ref{fig:maj2pt} show a numerical computation of the two-point function that bears out these conclusions.

Finally, let us estimate the height of the correlator plateau written down in \eqref{Gintt} in the cases where it does not vanish. This requires us to estimate the typical size of the matrix elements $|\psi_{nm}|^2 = \left|\langle n|\psi|m\rangle\right|^2$. For typical eigenstates $|n\rangle$ and $|m\rangle$, ETH predicts that the matrix elements will be of the same order as for random states, which would give $|\psi_{nm}|^2 = 1/L$. However, in our case, $|n\rangle$ and $|m\rangle$ are related by the action of the $P$ operator, $|n\rangle = P|m\rangle$, so we are actually considering a diagonal expectation value $|\psi_{nm}|^2 = |\langle m| \psi P|m\rangle|^2$. Then ETH instructs us to estimate this by replacing $|m\rangle$ with a random state.
One can check that this also gives $|\psi_{nm}|^2 \sim 1/L$.
The height of the plateau in the correlation function should then be of order $1/L$.

Notice that the spectral form factor plateau at $\beta=0$, given by \eqref{timeavgDeg}, is also of order $1/L$.
We therefore expect the correlator plateau and the spectral form factor plateau to be of the same order.
This holds true for the $N=18$ data, where the correlator plateau height is approximately $0.0075$, and the spectral form factor plateau is at approximately $0.004$.

\subsection{The ramp in more general theories}

One goal of this work is to evaluate how generic the ramp/plateau structure is in chaotic quantum field theories. In this section we ask whether it is plausible for this structure to appear in two-point functions of the form $G(t) = \langle O(t) O(0) \rangle$ in such theories. We will make two assumptions: that ETH holds for the theory, and that the late time behavior includes a ramp, as predicted by RMT.

If $O$ is a simple operator, we expect the two-point function to approach its time average $G_p$ plus fluctuations of order $e^{-S}\sim 1/L$. We would like to make sure that ramp behavior is consistent with this expectation.

The correlator can be written in the energy basis as
\begin{align}
  \langle O(t) O(0) \rangle &= 
  \frac{1}{Z(\beta)} \sum_{n} e^{- \beta E_n} |O_{nn}|^2 + 
  \frac{1}{Z(\beta)} \sum_{\begin{smallmatrix}
      n,m \\ E_n \ne E_m
    \end{smallmatrix}} e^{- \beta E_m} |O_{nm}|^2 e^{i(E_m -E_n)t} \ed \label{OO}
\end{align}
Here we assumed the spectrum is non-degenerate for simplicity.
The first sum in \eqref{OO}, coming from terms with $E_n = E_m$ in the double energy sum, exactly gives the plateau height $G_p$.
If the diagonal matrix elements $|O_{nn}|^2$ are of order unity then we find a plateau height $G_p \sim 1$ as discussed above.

The second sum in \eqref{OO} encodes correlations between different energy levels.
Beyond the dip time, it is responsible for the linear time dependence of the ramp.
ETH predicts that the off-diagonal matrix elements $|O_{nm}|^2$ are of order $1/L$ --- much smaller than the diagonal ones.
To get an estimate for the second sum we assume that these matrix elements can be treated as constant, $|O_{\rm off-diag.}|^2 \sim 1/L$.
The remaining sum then describes the ramp of the spectral form factor, sans the plateau contribution.
Altogether, the two-point function \eqref{OO} is given schematically by 
\begin{align}
  G(t) &\sim G_p + |O_{\rm off-diag.}|^2 \cdot
  Z(\beta) \cdot \left( \frac{t}{L^2} - \frac{1}{L} \right)
  \cr &\sim
  G_p + \frac{t}{L^2} - \frac{1}{L} \ed
\end{align}
Note that the $Z(\beta)$ factor in front of the parentheses is needed because the correlator is normalized differently than the spectral form factor. In writing the above expression, we are imagining that we are averaging over time somewhat, in order to supress fluctuations of $G(t)$ and get a smooth ramp. This type of averaging will be discussed further in section \ref{single}. In any case, the conclusion of this analysis is that $\langle O(t) O(0) \rangle - G_p \sim \frac{t}{L^2} - \frac{1}{L}$; the difference is suppressed by $1/L$ as expected.

\section{Single realization of random couplings}
\label{single}

\begin{figure}[h]
 \centering
 \includegraphics[width=0.6\textwidth]{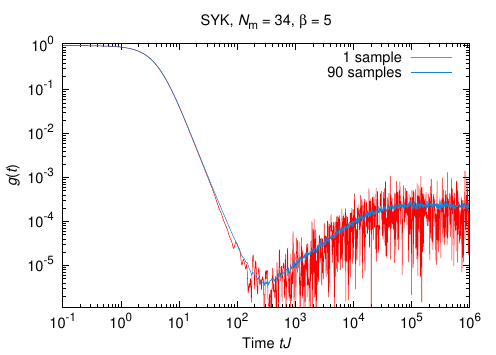}
 \caption{
 A single sample (red, erratic) of $g(t)$ is plotted together with the average of 90 samples (blue, smooth).}
 \label{fig:oneSample}
\end{figure}

It is important to ask whether the late time features of the spectral form factor (the dip, the ramp and the plateau) appear in ordinary chaotic quantum field theories without a disorder average.
As a first step towards addressing this question, in this section we consider the SYK spectral form factor $g(t;\beta)$ \eqref{g}, computed for a single realization of the random couplings $J_{ijkl}$.
Figure~\ref{fig:oneSample} compares the single sample result with the disorder averaged spectral form factor.
Before the dip time there is good agreement between the single sample and the averaged results.
This is consistent with our expectation that in the large $N$ limit and at early times, a typical sample should give a good approximation to the disorder-averaged spectral form factor.
We say that the spectral form factor is \textit{self averaging} at early times.

At late times, and in particular after the dip time, the spectral form factor is not self averaging \cite{prange1997spectral}.
This implies that at large $N$ a typical realization of the couplings does not give a result that approaches the disorder-averaged value.
In particular, at late times a typical realization exhibits large fluctuations as shown in Figure~\ref{fig:oneSample}.
We expect ordinary quantum field theories (with no disorder average) to have similar behavior.\footnote{We also expect the model recently discussed in \cite{Witten:2016iux} to behave similarly.}

Despite the large fluctuations, the underlying dip, ramp and plateau are still clearly visible.
These features can be made clear by averaging over a sliding time window of width $t_{\rm ave}$, smearing out the fluctuations.\footnote{Such time averaging and estimates compatible with ours have already been discussed in \cite{prange1997spectral}.}$^,$\footnote{Another possible way to reduce fluctuations in a CFT is to introduce a weak form of disorder averaging, by averaging slightly over the value of a marginal coupling.
  }
For this to work we need to be able to take  $t_{\rm ave}$ parametrically shorter than the length of the ramp, so that the features we are interested in will not get smeared out along with the fluctuations.
To estimate the required size of $t_{\rm ave}$, consider the auto-correlations in the random variable $|Z(\beta,t)|^2$,
\begin{align}
  h(t,dt;\beta) =
  \langle |Z(\beta,t)|^2 |Z(\beta,t+dt)|^2 \rangle -
  \langle |Z(\beta,t)|^2 \rangle \langle |Z(\beta,t+dt)|^2 \rangle \ed
  \label{autocorr}
\end{align}
Set $t$ to be a fixed time greater than the dip time.
At such fixed $t$ the autocorrelation $h(t,dt;\beta)$ decays with $dt$ with a typical time scale $t_{\rm decay}$.    After $t_{\rm decay}$ the signal is essentially uncorrelated.  
As we will see shortly, for $t$ on the plateau and at large $N$ and large $\beta$ we have $t_{\rm decay} \sim 1/\sqrt{N}$ .   For $t$ on the ramp $t_{\rm decay} \sim 1$ (We have suppressed the $\beta, J$ dependence here).  

Given $t_{\rm decay}$ we can estimate the minimal value of $t_{\rm ave}$ required to remove the fluctuations from the single-sample data.
We expect the fractional standard deviation (the ratio of the standard deviation to the mean) for such averaged data to behave like $\sqrt{\frac{t_{\rm decay}}{t_{\rm ave}}}$.
To have a curve with fluctuations smaller than, say, $1/N^2$ would require $t_{\rm ave}$ to be no greater than $N^4$, even if $t_{\rm decay}$ is as large as 1.  Such a value of $t_{\rm ave}$ is parametrically smaller than the length of the ramp, which is exponentially large in $N$.    
Therefore, at large $N$ the averaging width $t_{\rm ave}$ can be taken to be parametrically smaller than the length of the ramp.

Now we show that  on the plateau $t_{\rm decay} \sim 1/\sqrt{N}$ at large $N$ and large fixed $\beta$.
In the energy eigenbasis the autocorrelation function can be written as
\begin{align}
  h(t,dt;\beta) &=
  \sum_{k,l,m,n} \langle
  e^{-\beta (E_k + E_l + E_m + E_n)}
  e^{i t(E_k - E_l) + i (t+dt) (E_m - E_n)}
  \rangle
  \cr &\quad
  - \sum_{k,l,m,n} \langle
  e^{-\beta (E_k + E_l)} e^{i t(E_k - E_l)} \rangle
  \langle
  e^{-\beta (E_m + E_n)} e^{i (t+dt) (E_m - E_n)}
  \rangle \ed
\end{align}
Let  $t$ be greater than the plateau time $t_p$.
For such $t$, and when $dt$ is small, the first sum is dominated by terms which obey $E_k - E_l + E_m - E_n = 0$.
For a chaotic spectrum we generally expect only two solutions.
One solution, $E_k=E_l$, $E_m=E_n$, cancels with the disconnected part.
The second solution, $E_k = E_n$, $E_l = E_m$, gives the approximate answer
\begin{align}
  h(t,dt;\beta) \approx
  \sum_{m,n} \langle
  e^{-2 \beta (E_n + E_m)}
  e^{i dt (E_m - E_n)} \rangle = \langle Z(2\beta) \rangle^2 \cdot g(dt;2\beta) \ed
\end{align}
(Here we assumed that there are no degeneracies for simplicity.)
We find that at very late times $t$ the time dependence of the autocorrelation is given by the spectral form factor $g(dt;2\beta)$ at \textit{early} times.

At large $\beta$ and small $dt$, equation \eqref{slopelargeN} provides a good approximation to the spectral form factor, which decays as $g(dt;2\beta) \sim \exp \left( - cN dt^2 / (2\beta)^3 \right)$.
The typical decay time scales as $t_{\rm decay} \sim 1/\sqrt{N}$, as advertised above.
  After a time of order a few $\beta$ the exponential decay is replaced by a $1/(dt)^3$ power law decay.
  By this time the spectral form factor (and hence the autocorrelation $h$) is already exponentially suppressed.   A $1/(dt)^3$ power law is integrable so it does not alter our above estimate for the required time averaging window.
 
On the ramp the analysis is more subtle.  First we  make the plausible assumption that the  leading multipoint energy eigenvalue correlation functions at the exponentially small scales appropriate to the ramp are the same as the RMT correlators, up to $1/N^q$ corrections.   In GUE these correlators factorize into sums of products of  sine kernels \cite{mehta2004random}.   Then we can use a procedure like that leading to \eqref{ZZbargeneral} to show that for most of the ramp after a time $dt$ of order $1$ the autocorrelation function $h(dt)$ decays like $1/(dt)^4$.    For the earliest part of the ramp, $t < e^{N s_0}$,   $h(dt) \sim 1/(dt)^3$.     These power laws are integrable and so we estimate that on the ramp $t_{\rm decay} \sim 1$.   This means that $t_{\rm ave}$ can be chosen to make the error smaller than any power of $N$ and still leave exponentially many data points on the exponentially long ramp.   Numerics are not conclusive here, but do show a systematic decrease of error after time averaging.

\section{Conjecture about super-Yang-Mills}
\label{N4}

The above ideas make it possible to give a conjecture about the behavior of $g(t)$ and correlation functions $G(t)$ in the canonical $AdS$/CFT duality $AdS_5$/CFT$_4$ where $CFT_4$ is the $N=4$ SU(N) super Yang-Mills theory on $S^3$.    We will assume that the fine grained spectral statistics of this system are described by random matrix theory.   This seems highly plausible given  that this system at large `t Hooft coupling $\lambda$ is maximally chaotic, i.e., it saturates the chaos bound \cite{Maldacena:2015waa}.    We also assume that there are no new intervening nonperturbative time scales governing the behavior of $g(t)$ between the relatively short times governed by gravity and the very long times governed by random matrix theory.  A distinctive aspect of this system compared to SYK is that at very high temperatures $T$   the entropy becomes parametrically large and the plateau parametrically high relative to the dip.

There is no ensemble of Hamiltonians in this system so we want to describe the time averaged behavior of $Z(\beta, t)$ as discussed in the previous section.     We can relate this to the full density of states $\rho(E)$

\begin{equation}\label{symz}
Z(\beta, t) = \int_0^{\infty} dE \rho(E) e^{-(\beta +it) E} ~.
\end{equation}

At early times and large $N^2$ we evaluate \eqref{symz} by saddle point and use the bulk gravitational action to determine $\rho(E)$.   
 The initial behavior of $Z(\beta +it)$ should then  be given by analytically continuing the large euclidean $AdS$--Schwarzschild black hole action to complex $\beta$.   

In the following we use the results and follow the notation of \cite{Emparan:1999pm}. The black hole  metric  has warp factor $V(r) = 1- \mu/r^{n-2} +r^2/l^2$  where $\mu \sim G_n M$, $n+1 = D$ is the bulk spacetime dimension, and $l$ is the $AdS$ radius.     The horizon radius  $r_+$ is determined by $V(r_+) = 0.$    
 
 The inverse temperature is determined by finding the periodicity of time of the Euclidean signature metric, 
  \begin{equation}\label{betarplus}
 \beta = 4 \pi (l^2 r_+)/(n r_+^2 + (n-2) l^2) ~.
 \end{equation}
 
 The action $I, ~     (Z = e^{-I})$ is given by  
\begin{equation}\label{gravaction_BH}
  I_{BH}= \frac{C}{G_N} \beta\left( -r_+^n + r_+^{n-2} l^2
    + \frac{3}{8} l^4\right) \ed
\end{equation}
Here and below $C$ is a positive constant and $G_N \sim \frac{1}{N^2}$.   The $ \frac{3}{8} l^4 $ term is specific to $n=4, D=5$. These Casimir energy type terms are missed without thinking about holographic renormalization \cite{Emparan:1999pm}. The thermal $AdS$ action in this scheme is
\begin{equation}\label{gravaction_AdS}
I_{AdS} = \frac{C}{G_N} \beta \cdot \frac{3}{8} l^4 \ed
\end{equation}
We find it convenient to subtract the ground state energy, and study \eqref{symz} via $e^{-I}$ where $I= I_{BH} -I_{AdS}$. In other words, we do not include the Casimir term.

Now we analytically continue.  As $\beta \rightarrow \beta +it$, $r_+$ becomes complex.     For small real $\beta$, $r_+ \sim 1/\beta$.   Adding a small positive imaginary part to $\beta$ corresponds to adding a small negative imaginary part to $r_+$.    At large $t$,  $r_+ \rightarrow   -i \sqrt{\frac{n-2}{n}} l $. 

At  $t=0$,  $Z \sim e^{c N^2/\beta^3}$ which for small $\beta$ is very large.  This reflects the very large entropy at high temperatures. (Here and below $c$ denotes positive constants of order one.)  As $t$ is increased and $\beta \rightarrow \beta+ it$,  $|Z|^2$ drops very quickly.   At $t \sim \beta$, $|Z|^2$ becomes less than one.   Then another saddle, the thermal $AdS$ solution, dominates.   Including the one loop determinant around this saddle representing a gas of gravitons we find $|Z|^2 \sim e^{c/\beta^3}$, an $N$-independent much smaller value.\footnote{To be precise, there are  of order $\frac{1}{\beta^3}$ weakly interacting gravitons of $AdS$ energy  and so $Z(\beta +it)$ oscillates with $AdS$ frequency.}

But this is not the whole story.   As $t$ increases to $AdS$ scale  $|Z|^2$ increases again, eventually becoming of order $|Z|^2 \sim e^{c N^2}$ (with  no $\frac{1}{\beta^3}$ enhancement).   This apparently dominates over the thermal $AdS$ again.   

But there is another wrinkle.  As $t$ becomes large the solutions of $V(r_+) = 0$ coalesce.  This causes the fluctuation corrections to  the saddle point in \eqref{symz} to behave like $\frac{t}{N^2}$, becoming large at $t \sim N^2$.   Taken at face value these large corrections invalidate the saddle point analysis for times larger than this.

Other saddle points could be relevant here.  For example, at high temperature, small $\beta$, there is a 10D small  Schwarzschild black hole saddle point with $r_+ \ll l$.   Using the $n=9$ version of the above formulas gives an initial $|Z|^2 \ll 1$, evolving at $t \sim \beta$ to $|Z|^2 \sim e^{c N^2}$ and then rapidly decreasing to $|Z|^2 \ll 1$ again.   
But at $t \sim l$, $r_+$ becomes of order $l$ and $AdS$ corrections become important.  
A more careful analysis would be required.

Although it is never thermodynamically dominant,
the recent analysis of \cite{Dias:2016xgx} indicates that there is a Gregory-Laflamme-type 5D to localized 10D transition in the space of saddles.  At first glance this could produce a singularity in $\rho(E)$ leading to a slow long-time falloff.  If this transition is caused by a single mode becoming tachyonic then it produces a branch point singularity in $Z$ which presumably can be analytically continued around.   If there is a more serious kind of large $N$ transition it may produce a more extreme form of singularity.   In any event, at large but finite $N$ this feature will be smoothed out, so we do not expect it to produce significant long-time effects past times polynomial in $N$.    

Although this analysis is far from conclusive\footnote{As an example of the subtleties here,   for $n=5, D = 6$  the dominant saddle causes $ | Z|^2$ to diverge as $t \rightarrow \infty$. This is inconsistent so presumably this saddle eventually leaves the  integration contour. In general we have not attempted to decide which saddles are on or off the steepest descent contour.} it does seem like the most plausible values of $|Z|^2$ in the slope region leading up to the dip have $N$ scaling $|Z|^2 \sim 1$ or $|Z|^2 \sim e^{cN^2}$ .   We will assume these values and compute the dip by matching onto the ramp, to which we now turn.

\subsection{The ramp in SYM}

SYM at large `t Hooft coupling $\lambda$ is maximally chaotic according to the out of time order correlator diagnostic, so it is plausible to conjecture that its fine grained eigenvalue statistics are described by random matrix theory.   The relevant ensemble will be determined by the symmetries of the system.   
For simplicity let us imagine that a nonzero $\theta$ term is present to break the $T$ symmetry.  Then we expect GUE statistics.\footnote{We thank Alex Maloney for this observation.} For simplicity, we will discuss the case of a high temperature state, where $\beta$ is small compared to the spatial $S^3$ radius.

We can outline a simple expectation for the ramp behavior using the formula (\ref{ZZbargeneral}). We interpret the expectation value $\langle \cdot \rangle$ as denoting a time average rather than a disorder average, as discussed in section \ref{single}. The procedure is equivalent to ``unfolding'' the spectrum, analyzing the ramp and plateau for each narrow energy band, and then adding them up together. First, we study the ramp at reasonably late time, $t>e^{\# N^2}$, where the relevant energies will be high enough that we can use planar SYM formulas for $S(E)$:
\begin{align}
  \log Z &= S-E/T = c_0 N^2 T^3 \ec \\
  E &= 3 c_0 N^2 T^4 \ec \\
  S  &= 4 c_0N^2 T^3  = \frac{4}{3^{3/4}} c_0^{1/4} N^{1/2} E^{3/4}.
\end{align}

At large $N$ the integral in (\ref{ZZbargeneral}) will be sharply peaked, and the band that makes the largest contribution at time $t$ will be the band which is just reaching the plateau at time $t$. That is, $S(E) = \log t$.  Using the above equations we then have
 \begin{equation}\label{newramp}
 g_{\rm ramp}(t) =\frac{t}{Z(\beta)^2} \exp\left[-\frac{3\beta}{2^{5/3}c_0^{1/3}}\frac{(\log t)^{4/3}}{N^{2/3}}\right].
 \end{equation}
 The growth is somewhat slower than linear, and the ramp joins the plateau at time $t_p = e^{S(2\beta)} = e^{4c_0N^2/(2\beta)^3}$ where the derivative of (\ref{newramp}) vanishes.
 
 To understand the dip time $t_d$ we need to work out the behavior of the ramp at earlier times. It is possible that weak interactions in the $AdS$ gas could lead to a small ramp, but we focus our attention on the region where the ramp would be associated to black hole states. The smallest black holes that dominate the microcanonical ensemble are determined by microscopic parameters, as discussed by \cite{ Susskind:1993ws, Horowitz:1999uv, Horowitz:1996nw}, but in fact these black holes give contributions to the ramp that are smaller than the slope contribution to $g(t)$. To see this, we consider 10D Schwarzschild black holes of mass $E$ with Schwarzschild radius $r_s$  much smaller than the $AdS$ radius $l$ where (in the remainder of this section we suppress numerical factors)
 \begin{align*}
 E &= r_s^7/G_N \ec \\
 S &= r_s^8/G_N \ed
 \end{align*}
 Here $G_N = l^8/N^2$ is in 10D. The contribution of such black holes to the ramp would be
 \be\label{smallramp}
g_{ramp}(t)\sim\frac{1}{Z(\beta)^2}\int dr_s \, e^{-N^2 \beta r_s^7/l^8}\text{min}\left\{t,e^{N^2 r_s^8/l^8}\right\} \approx \frac{t}{Z(\beta)^2}\exp\left[-N^{1/4}\frac{\beta}{l}(\log t)^{7/8}\right],
 \ee
where we used that the integral is dominated by the value of $r_s$ such that $t = e^{N^2 r_s^8/l^8}$. The dip time is the first time such that this contribution is larger than the contribution of the slope. The expected slope contribution is no smaller than $\frac{1}{Z(\beta)^2}$. Eq.~(\ref{smallramp}) first exceeds this value when the dominant value of $r_s$ is $r_s = \beta$, or equivalently at a time
 \be
t_d = e^{N^2 \beta^8/l^8}.
 \ee
So we see that the first black holes that are relevant are small, with $r_s\sim\beta$, but not microscopic. The associated dip time is exponential, but with a parametrically small coefficient at high temperature. Note that this conclusion is rather sensitive to the long-time behavior of the slope, so this identification of the dip time is tentative. For example, if we have $g_{slope} = \frac{1}{Z(\beta)^2}e^{c N^2}$ instead, then we would expect $t_d \sim e^{c' N^2}$ for an order one $c'$. Either way, at high temperature we have a large hierarchy between the dip time and the plateau time $t_p\sim e^{4 c_0 N^2/(2\beta)^3}$, leading to a parametrically long ramp.\footnote{In fact here the hierarchy is more dramatic than in SYK because the plateau can be made arbitrarily high by increasing the black hole temperature.}  The early time behavior of  $AdS_3$/CFT$_2$ is under greater analytic control and has been analyzed in \cite{Dyer:2016pou} .  There is also an exponential hierarchy, although not as large,  in this system.

It would be interesting to consider observables that probe the ramp during earlier times where microscopic black holes are relevant. One possibility would be to directly consider a microcanonical partition function that selects this part of the spectrum.

 There is a subtlety in these estimates.   SYM and many other theories have global symmetries (like the SO(6) R symmetry).   We expect the spectrum within each sector to have chaotic RMT correlations, but the different sectors would be essentially uncorrelated.   We expect the number of thermodynamically significant sectors at fixed $\beta$ to be at most  polynomial in the entropy $S$.   If we denote the separate sectors by indices $a, b$ we can write $g(t) = \sum_{a, b} g_{ab}(t) $ where $g_{ab}$ contains the sum over energies in the fixed sectors $a, b$.   The diagonal terms in this sum contribute as usual to the ramp and plateau; the off-diagonal terms have vanishing contribution  at late time and large $S$.   So the overall heights of the ramp and plateau are suppressed by polynomials in $S$.   This effect is subleading to the exponential effects we are interested in and so we ignore it.    We have confirmed these ideas in the Dirac SYK model which contains a U(1) charge.
 
\section{Discussion}
\label{discussion}

In this paper we have argued that the late time behavior  of horizon fluctuations in large $AdS$ black holes is governed by random matrix dynamics.   Our main tool was the SYK model, which we used as a simple model of a black hole, adequate for such qualitative questions.

Using numerical techniques we established random matrix behavior at late times.   We were able to determine the early time behavior precisely in the double scaling limit.  This enabled us to give a plausible estimate for the dip time by computing the intersection of these two curves.\footnote{As noted earlier there could be new phenomena at early times, like spin-glass behavior, or $1/N^{q}$ effects absent in the double scaling limit. It seems quite plausible these would cause the slope to decay faster and so make the dip time earlier.}   The dip time is exponentially late, and the ramp region, controlled by long-range spectral rigidity, is exponentially long, stretching until the asymptotic plateau behavior sets in.

It will be useful to have analytic insight into the ramp behavior in the SYK model.  In Appendix \ref{Nq} we make some preliminary remarks about the origin of the $e^{-2S}$ scale of the height of the ramp in this model.

We used these ideas to formulate a conjecture about more general large $AdS$ black holes, like those dual to 4D SYM theory.   Here we rely on the widely accepted intuition that the fine grained structure of energy levels in chaotic systems is described by random matrix theory.  We then estimated the time at which the ramp appears by making a provisional estimate of the analytically continued 5D $AdS$-Schwarzschild black hole metric.   Again we find an exponential hierarchy of scales.\footnote{In fact here the hierarchy is more dramatic because the plateau can be made arbitrarily high by increasing the black hole temperature.}  The early time behavior of  $AdS_3$/CFT$_2$ is under greater analytic control and has been analyzed in \cite{Dyer:2016pou} .  There is also an exponential hierarchy in this system.

In all of these situations the dip time does not signal a new physical phenomenon\footnote{If the spin glass or $1/N^q$ possibilities are present then there is a new physical phenomenon at the dip.}  -- it is just the time when the ramp becomes larger in size than the slope.  To understand the new physics of the ramp
it would be interesting to follow it  ``underneath" the slope to see what happens at early times.  For instance in SYM one would start accessing regions controlled by string scale black holes, and eventually the chaotic graviton gas.   To do this it might be useful to use a more refined probe than $g(t)$.  

A more indirect strategy would be to look for precursors of the ramp starting from short times.   Do the individual terms in the $1/N$ expansion get large as time is increased?\footnote{The disconnected partition function $g_d(t)$ provides an example of this.   As discussed above, the gaussian fluctuations of the edge of the eigenvalue distribution produce a gaussian falloff in $g_d(t)$.  These are signaled by a series of terms of the form $(t^2/N^{q-2})^k$.  This softening cancels out in the time dependence of $g(t)$. }    Or is there just a factorial growth of coefficients signaling an asymptotic expansion with an exponentially small error sufficient to accommodate the ramp and plateau signals?     Knowing this would be helpful in looking for signals in SYM of these phenomena.

To understand the SYM situation better it would be useful to understand more about the averaging procedures that are available.    Averaging over time windows has been discussed in Section~\ref{single}.  But perhaps one could take an ensemble of SYM theories with slightly different parameters.    This possibility may be easier to implement in calculations.

Another set of ideas  that might be useful are developments in the theory of sparse random matrices.  From this point of view the SYK model is a certain type of sparse random matrix with correlated randomness in the entries.  Insights have emerged  \cite{rodgers1988density,2014arXiv1403.1114K, erdHos2012universality, erdHos2014phase} about the universality of dense random matrix behavior in the fine grained eigenvalue statistics of various types of sparse matrices.  These might give clues about the SYK model, and more general contexts.  This is a question we would like to return to in future research.

Perhaps the central question this work raises is the nature of the  bulk interpretation of the random matrix behavior.   The disorder averaged SYK model can be rewritten exactly in terms of the bilocal collective fields $G(t, t'), \Sigma(t, t')$.     For $g(t)$ one needs two copies (``replicas") of the fermions and so the $G_{\alpha\beta}, \Sigma_{\alpha\beta}$ fields carry replica indices  $\alpha, \beta = 1,2$.
An appropriate contour can be chosen so the functional integral over $G_{\alpha\beta}(t,t'), \Sigma_{\alpha\beta}(t,t')$ is nonperturbatively well-defined, as discussed in Appendix \ref{toyapp}.   This functional integral is a rough proxy for a bulk description, because it involves $O(N)$ singlet objects and in some rough way bulk fields should be able to be reconstructed from the bilocal singlet fields.   

This functional integral must contain the ramp and plateau behavior -- the question is how.   We cannot yet answer this question -- it will continue to be a focus of our research.  Here we will just make some preliminary comments. 

The coefficient of the $G, \Sigma$ action includes $N$, so new saddle points are a natural mechanism for such $e^{-N}$ effects.  For $q=2$, as explained in Appendix \ref{qequal2},  quenched correlators do seem to be described by sums over new saddle points with appropriate fluctuation corrections.    Here the ramp is a perturbative $1/N^2$ effect and the plateau is an $e^{-N}$ effect.

For the interacting case $q >2$ the situation is qualitatively different.   Here the interplay between $L$ and $N$ discussed in Section \ref{rampRMT}  becomes crucial.      The ramp is a $1/L^2$ effect, which means an $e^{-N}$ effect.
In Appendix~\ref{Gramp} we point out obstacles to a possible single saddle point explanation of the ramp in the correlator $G(t)$.
But various auxiliary quantities like the $f_k$ discussed in Appendix \ref{Nq}   can be computed by saddle point, giving the desired $2^{-N}$ value for large $k$.   It is unclear whether this has anything to do with an actual saddle point description of the ramp involving a sum over many saddle points.  

The $N$ mod 8 ``eightfold way'' pattern noted in \cite{PhysRevB.83.075103,You:2016ldz,Fu:2016yrv} must have an explanation in the $G, \Sigma $ integral.    In some ways it seems analogous to the behavior of the Haldane spin chain \cite{haldane} as the spin varies from integer to half integer.  There the explanation in the continuum is a topological term in the action.    That would be a natural guess here, and the question is what topology is being probed.   As an initial  step it will be  important to find the origin of this effect in the moment calculations discussed in Appendix \ref{Nq}.

The origin of the plateau in the $G, \Sigma$ integral is another mystery.   After continuing to imaginary energy this is an effect of order $\exp{(-L)}$ which is of order $\exp{(-e^{N})}$.  This is an unusually small nonperturbative effect, the size of the error in an asymptotic series of multi-instanton corrections.    A more natural way to explain these effects would be a mapping from $G, \Sigma$ to new effective random matrix degrees of freedom with effective coupling $1/L$ whose dynamics would give the plateau as a standard Andreev-Altshuler instanton nonperturbative effect \cite{andreev1995spectral, kamenevmezard}.   This map would be related to reconstituting the fermions from the collective fields.\footnote{Some ways in which fermionic properties are coded into $G, \Sigma$ are discussed in Appendix \ref{toyapp}.}  This is a challenging proposition but the SYK model provides the most concrete arena known in which to explore it.  

\section*{Note added}
Minor plotting errors were present in  versions v1 and v2, and in the published version of this manuscript \cite{Cotler2017}. 
These errors do not affect any of the qualitative or quantitative statements made in the paper.  We have corrected the errors (listed below) in this version and submitted an erratum for \cite{Cotler2017}.   
\begin{itemize}
\item Figures
\ref{fig:g-SYK},
\ref{fig:g-RMT}, 
\ref{fig:g-majorana-beta0}, \ref{fig:largeNg},
\ref{fig:oneSample},
and \ref{fig:f-ggcgdt}
displayed $g(t, \beta)$ (and $g_c(t, \beta)$ and $g_d(t, \beta)$ in Fig.~\ref{fig:f-ggcgdt})
obtained by using $Z(\beta, t_0)$ instead of $Z(\beta) = Z(\beta, t=0)$ as the normalizing factor $Z(\beta)$ in \eqref{g}
(and in \eqref{gd} and \eqref{gc} for Fig.~\ref{fig:f-ggcgdt}), in which $t_0 = 0.1 J^{-1}$ is the smallest value of $t$ in the plots.   The effect of this is a small ($\sim 1\%$) uniform vertical shift of the graphs.  This is barely visible on the figures.

\item Figure~\ref{fig:g-RMT} was originally produced by treating two samples of the eigenvalues of $2^{12}$-dimensional matrices as a single sample of a $2^{13}$-dimensional matrix.

\item We used different time steps in Figure 10 for the single sample and 90 sample data  in the original versions.
We have taken advantage of this correction to use a different, more appropriate, scale on the plot.

\item 
Only 914 samples were used for both Figs.~\ref{fig:g-majorana-beta0} and Fig.~\ref{fig:f-ggcgdt}, instead of the  1000 stated.
Also, the number of samples for $N=16$ was 1 000 000 in Fig.~\ref{fig:g-majorana-beta0} and 1 200 000 in Fig.~\ref{fig:f-ggcgdt}.
In this revised version, 1 200 000 samples for $N=16$ and 914 samples for $N=30$ have been used for both figures.
\end{itemize}

\section*{Acknowledgements}
The authors thank Tom Banks, Ethan Dyer, Alexei Kitaev, Juan Maldacena, Alex Maloney, Dan Roberts, Lenny Susskind and Edward Witten for discussions.

This work was partially supported by JSPS KAKENHI Grant Numbers JP25287046 (M.H.), JP15H05855 (M.T.) and JP26870284 (M.T.).  J.C. is supported by the Fannie and John Hertz Foundation and the Stanford Graduate Fellowship program.
G. G. is supported by a grant from the John Templeton Foundation. The opinions expressed in this publication are those of the authors and do not necessarily reflect the views of the John Templeton Foundation.
S. S. is supported by NSF grant  PHY-1316699.  D.S. is supported by the Simons Foundation grant 385600.   Preliminary versions of this work were presented at the Yukawa Institute for Theoretical Physics Workshop (June 9, 2016) and  at ``Natifest" at the Institute for Advanced Study (September 17, 2016).  We thank these institutions for their hospitality.  In accordance with institutional policy the data used in the preparation of this paper is available to other scientists on request.

\appendix

\section{Particle-hole symmetry of SYK}
\label{P}

In the Dirac description \eqref{chiDirac} the Hamiltonian has conserved charge parity, where the charge (fermion number) operator is $\hat{Q} = \sum_i \cd_i c_i$.
The Hamiltonian \eqref{Hmajorana} has two sectors for charge parity even and odd.

The theory also has a particle-hole symmetry under the operator \cite{PhysRevB.83.075103,You:2016ldz,Fu:2016yrv}
\begin{align}
  P = K \prod_{i=1}^{N_d} (\cd_i + c_i)
\end{align}
where $K$ is the anti-linear operator that takes $z \to \bar{z}$, $z \in \bC$  (here we choose $c_i,\cd_i$ to be real with respect to $K$).
One can check that
\begin{align}
  P^2 =(-1)^{\frac{N_d(N_d-1)}{2}}= 
  (-1)^{\left\lfloor N_d/2 \right\rfloor} =
  \left\{
  \begin{matrix}
    +1 & ,\, N_d~\mathrm{mod}~4 = 0 \\
    +1 & ,\, N_d~\mathrm{mod}~4 = 1 \\
    -1 & ,\, N_d~\mathrm{mod}~4 = 2 \\
    -1 & ,\, N_d~\mathrm{mod}~4 = 3
  \end{matrix}
  \right. \ed
\end{align}
The action on the fermions is given by
\begin{gather}
  P c_i P = \eta \cd_i \ecq
  P \cd_i P = \eta c_i \quad \Rightarrow \quad
  P \psi_a P = \eta \psi_a \ec\label{actionOnFermions}
\end{gather}
where
\begin{align}
  \eta = (-1)^{N_d-1} P^2 = 
  (-1)^{\left\lfloor \frac{3N_d}{2}-1 \right\rfloor} \ed
\end{align}

One can now check that $P$ is a symmetry,
\begin{align}
  [H,P] = 0 \ed
\end{align}
For some values of $N_d$ this leads to a degeneracy in the spectrum.
$P$ maps a state with fermion number $Q$ to $N_d-Q$ (in our convention the Fock space vacuum has fermion number 0).
\begin{enumerate}
  \item If $N_d = N/2$ is odd then $P$ maps the even and odd charge parity sectors to each other, and so the two sectors are degenerate.
  \item If $N_d = N/2$ is even then $P$ maps each charge parity sector to itself.
    \begin{enumerate}
      \item If $(N_d~\mathrm{mod}~4) = 2$ then $P^2=-1$. 
        Since $P$ is both anti-linear and obeys $P^2=-1$ it cannot map energy eigenstates states to themselves, and we have double degeneracy within each sector.
      \item If $(N_d~\mathrm{mod}~4) = 0$ then $P^2=1$.
        In this case there is no protected degeneracy.
    \end{enumerate}
\end{enumerate}
Therefore, for $(N~\mathrm{mod}~8) \ne 0$ there is 2-fold degeneracy, while for $(N~\mathrm{mod}~8) = 0$ there is no protected degeneracy.

\section{The double-scaled SYK theory}
\label{doubleScaledSYK}
In this appendix we compute the disorder-averaged spectrum of the SYK theory in the double-scaled limit
\be
N\rightarrow \infty , \hspace{20pt} q\rightarrow \infty, \hspace{20pt} \lambda = \frac{q^2}{N} = \text{fixed}.
\ee
The computation is a small modification of the analysis by Erd\H{o}s and Schr\"{o}der in \cite{erdHos2014phase} for closely related systems (composed of Pauli matrices with random couplings instead of Majorana operators).

The argument goes as follows: first, we compute the moments $\tr H^k$. Then, we appeal to a combinatoric result in \cite{ismail1987combinatorics} to get the distribution for which these are the moments. 

First, we discuss the computation of the moments. We would like to evaluate
\be
\langle \tr H^{k}\rangle_J
\ee
for $k$ even. We evaluate the $J$ integral by Wick contractions. This involves pairing up the various terms in different $H$ factors and contracting the flavor indices of the fermions that appear in the pair. If all of the Wick-contracted pairs were adjacent in the product, we could evaluate each pair as $\frac{1}{2^q}$, since $\psi_i \psi_i = \frac{1}{2}$. Taking the product over the pairs and summing over the possible fermion flavors that can occur in each pair, we get
\be
\frac{\tr H^k_{\text{assuming all pairs next to each other}}}{\tr 1} = \left[\frac{\langle J^2\rangle}{2^{q}}{N \choose q}\right]^{k/2} = \left(\frac{\mathcal{J}^2}{2\lambda e^{\lambda/2}}\right)^{k/2}.\label{assu}
\ee
where $\mathcal{J}$ is defined by $\langle J_{i_1...i_q}^2\rangle = \frac{2^{q-1}}{q}\frac{\mathcal{J}^2(q-1)!}{N^{q-2}}$ \cite{Maldacena:2016hyu}. Now, of course we also have to consider cases where Wick-contracted pairs are not adjacent. The procedure is to commute the terms past each other until the contracted pairs are adjacent or nested, so that Wick-contraction lines do not cross.

Let's consider what happens when we move one product of fermions past another. Notice that
\be
\left[\psi_{a_1}...\psi_{a_q}\right]\left[\psi_{b_1}...\psi_{b_q}\right] = (-1)^{\text{\# fermions in common}}\left[\psi_{b_1}...\psi_{b_q}\right]\left[\psi_{a_1}...\psi_{a_q}\right].
\ee
The important feature of the limit where we hold $q^2/N$ fixed is that the expected number of fermions in common stays of order one in this limit. More precisely, the number is Poisson distributed, with distribution
\be
P(m\text{ fermions in common}) = \frac{\lambda^{m}}{m!}e^{-\lambda},\hspace{20pt} \lambda = \frac{q^2}{N}.
\ee
Now, in principle, things will get complicated because we have to consider the possibility that the same fermions that are shared between two copies of the Hamiltonian containing $\psi_{a_1}\cdots\psi_{a_q}$ and $\psi_{b_1}\cdots\psi_{b_q}$ might also be shared with a third copy containing $\psi_{c_1}\cdots\psi_{c_q}$. Or, more generally, that the number of such terms might be correlated. However, the probability is proportional to $1/N$, without a $q^2$ enhancement, so we ignore it in the double-scaled limit. This is the key point that makes it possible to solve.

So now, each time we have to commute a product of $q$ fermions past each other, we get a factor
\be
\sum_{m = 0}^\infty (-1)^m P(m\text{ fermions in common}) = e^{-2\lambda}.
\ee
Notice that this step differs somewhat from the case considered by \cite{erdHos2014phase}. Specifically, this is where the fact that we have Majoranas instead of spins is relevant. In the spin case, the analogous sum gives $e^{-4\lambda/3}$. Anyhow, doing this sum independently for each set of fermions that we need to commute past each other, we can now correct the expression (\ref{assu}), and we find
\begin{align}
\frac{\tr H^k}{\tr 1} = \sum_{\text{Wick pairings}} \left(\frac{ \mathcal{J}^2}{2\lambda e^{\lambda/2}}\right)^{k/2} e^{-2\lambda \ \text{cross}(\text{pairing})}.
\end{align}
Here $\text{cross}()$ gives the number of commutations required to get the pairs arranged in a way so that they are all adjacent or nested. We can describe this target situation by saying that lines connecting the Wick pairs will not cross. Then $\text{cross}()$ is just the initial number of crossings of Wick contraction lines. 

The final step is to notice \cite{erdHos2014phase} that the distribution with these moments is known \cite{ismail1987combinatorics}. It is related to the integration measure for the $Q$-Hermite polynomials, with $Q = e^{-\lambda}$. The distribution is given by
\begin{align}
\rho(E) = \frac{\mathcal{N}}{\sqrt{1 - a^2}}\prod_{n=0}^\infty\left(1 - \frac{a^2}{\cosh^2(n\lambda)}\right), \hspace{20pt} a^2\equiv \frac{\lambda e^{\lambda/2}(1{-}e^{-2\lambda})}{2}\frac{E^2}{\mathcal{J}^2}.
\end{align}
for $|a|<1$ and zero otherwise. The normalization factor can be determined from the constraint that the total number of states is $2^{N/2}$.

It is convenient to rewrite $\rho$ as follows
\begin{align}
\log \frac{\rho(E)}{\mathcal{N}}
&= \frac{1}{2}\sum_{n = -\infty}^\infty \log\left(1 - \frac{a^2}{\cosh^2 (n \lambda)}\right)\\
&= \frac{1}{2}\sum_{k= -\infty}^\infty \int_{-\infty}^\infty dn\, e^{-2\pi i k n}\log\left(1 - \frac{a^2}{\cosh^2 (n\lambda)}\right)\\
&= -\frac{1}{\lambda}\left(\arcsin a\right)^2  + \sum_{k\ge 1}\frac{1 - \cosh\left[\frac{k\pi}{\lambda}\left(\pi - 2 \text{arccos}\, a\right)\right] }{k\sinh\frac{k \pi^2}{\lambda}}.\label{sumn}
\end{align}
In the second line we used the Poisson resummation formula. In the last line we did the $n$ integral by contour integration, summing over a geometric series of cuts of finite length along the imaginary $n$ axis. This formula is now in a convenient form for discussing the behavior at small $\lambda$.

For example, if we take $\lambda \rightarrow 0$ with $a$ fixed, the first term dominates, and exactly reproduces the large $q$ thermodynamics computed in \cite{Maldacena:2016hyu}.

Our primary goal is to use this to evaluate the partition function of the Schwarzian theory, so we take a further ``triple-scaled'' limit 
\be
\lambda \rightarrow 0, \hspace{20pt} \frac{E-E_0}{\mathcal{J}}\rightarrow 0, \hspace{20pt} z\equiv \frac{(E-E_0)}{\lambda \mathcal{J}} = \text{fixed}.
\ee
In this limit, we approximate $a = -1 + z\lambda^2 + O(\lambda^3)$ and we have
\begin{align}
\frac{\text{arccos}(a)}{\lambda} = \sqrt{2z} + O(\lambda), \hspace{20pt} \frac{(\arcsin(a))^2}{\lambda} = \frac{\pi^2}{4\lambda} - \pi\sqrt{2z} + O(\lambda) \ec\\ \cr
\sum_{k\ge 1}\frac{1 - \cosh\left[\frac{k\pi}{\lambda}\left(\pi - 2 \text{arccos}\, a\right)\right] }{k\sinh\frac{k \pi^2}{\lambda}} = \log\left(1 - e^{-2\pi\sqrt{2z}+O(\lambda)}\right) + O(e^{-\pi^2/\lambda}).
\end{align}
We conclude that in the triple-scaled limit we have
\be\label{sinhtriple}
\rho(E) = 2\mathcal{N}e^{-\frac{\pi^2}{4\lambda}}\sinh\left(\pi\sqrt{2z}\right), \hspace{20pt} z = \frac{(E - E_0)}{\lambda\mathcal{J}}, \hspace{20pt} \lambda = \frac{q^2}{N}.
\ee
One can check that for small $\lambda$ the normalization factor is $\mathcal{N} = \frac{2^{ N/2}}{\mathcal{J}}\sqrt{\frac{\lambda}{\pi}}$, which leads to
\be
Z(\beta) = \int dE \rho(E)e^{-\beta E} = e^{-\beta E_0 + S_0}\frac{\sqrt{2}\pi}{(\beta\mathcal{J})^{3/2}}\exp\left(\frac{\pi^2}{2\lambda\beta\mathcal{J}}\right).
\ee
where $E_0 = -\frac{\mathcal{J}}{\lambda}$ and $S_0 = N\frac{\log(2)}{2} - \frac{\pi^2}{4\lambda}$.
This agrees with the 1-loop calculation of \cite{Maldacena:2016hyu}, but here we conclude that it is the exact answer in the triple-scaled limit that isolates the Schwarzian.

Finally, we will mention that there is another way to analyze the double-scaled limit, starting from the $G,\Sigma$ action for the disorder-averaged partition function:
\be
-I = \frac{N}{2}\log\det(\partial_\tau - \Sigma) - \frac{N}{2}\int d\tau_1 d\tau_2\left[\Sigma G - \frac{\mathcal{J}^2}{2 q^2}(2G)^q\right].
\ee
To take the double-scaled limit, we write
\be
\Sigma(\tau_1,\tau_2) = \frac{\sigma(\tau_1,\tau_2)}{q}, \hspace{20pt} G(\tau_1,\tau_2) = \frac{\sgn(\tau_{12})}{2}\left(1 + \frac{g(\tau_1,\tau_2)}{q}\right).
\ee
where now $g(\tau_1,\tau_2)$ is a symmetric function of its two arguments that is constrained to vanish when they coincide. The action in the double-scaled limit is
\begin{align}
-I &= \frac{N}{4q^2}\Big[-\int d\tau_1..d\tau_4 \frac{\sgn(\tau_{12})}{2}\sigma(\tau_2,\tau_3)\frac{\sgn(\tau_{34})}{2}\sigma(\tau_4,\tau_1) \\
&\hspace{20pt}+ \int d\tau_1d\tau_2\left(\mathcal{J}^2 e^{g(\tau_1,\tau_2)}-\sgn(\tau_{12})\sigma(\tau_1,\tau_2)g(\tau_1,\tau_2) \right)\Big].
\end{align}
Notice that $\sigma$ appears quadratically, so we can integrate it out exactly. We get
\be
-I = \frac{N}{4 q^2}\int d\tau_1d\tau_2 \left[\mathcal{J}^2e^{g(\tau_1,\tau_2)} - \frac{1}{4}\partial_{\tau_1}g(\tau_1,\tau_2)\partial_{\tau_2}g(\tau_1,\tau_2)\right],
\ee
which has the form of a Liouville action on a Lorentzian space. One can analyze this theory by studying perturbation theory in $\mathcal{J}^2$. This is equivalent to computing moments as in the Erdos-Schroder analysis. Note that at each order in $\mathcal{J}^2$ we have a simple Gaussian integral.

\section{A toy $G,\Sigma$ integral}
\label{toyapp}
In the main text, we asserted that $G,\Sigma$ give a nonperturbatively exact formulation of the disorder-averaged SYK model. In this appendix, we discuss a toy model for the $G,\Sigma$ path integral. We discuss the contour of integration and saddle points, and we see how Grassmann behavior can arise from bosonic variables.

The example that we will discuss can be thought of as the SYK Grassmann path integral on a space where we replace the time dimension by two points, labeled $1$ and $2$. Then the fermion variables are $\psi_i(1),\psi_i(2)$ where $i = 1,...,N$. Concretely, the integral we consider for fixed disorder is
\begin{align}
Z = \int d^N\psi(1)d^N\psi(2) e^{\sum_{i}\psi_i(1)\psi_i(2) + \sum_{i_1<...<i_q}J_{i_1...i_q}[\psi_{i_1}(1)...\psi_{i_q}(1)+\psi_{i_1}(2)...\psi_{i_q}(2)]}.
\end{align}
The average over couplings gives
\begin{align}
\langle Z\rangle &= \int d^N\psi(1)d^N\psi(2) e^{\sum_i \psi_i(1)\psi_i(2) + \frac{(q-1)!J^2}{N^{q-1}}\sum_{i_1<...<i_q}\psi_{i_1}(1)\psi_{i_1}(2)...\psi_{i_q}(1)\psi_{i_q}(2)}\\
&= \int d^N\psi(1)d^N\psi(2) e^{\sum_i \psi_i(1)\psi_i(2) + \frac{J^2}{qN^{q-1}}[\sum_i\psi_i(1)\psi_i(2)]^q}.\label{starthere}
\end{align}
We can write this as a $G,\Sigma$ integral by the standard manipulation: we introduce a variable $\sigma$ that is a Lagrange multiplier that sets $g = \frac{1}{N}\sum_{i}\psi_i(1)\psi_i(2)$. This leads to the expression
\begin{align}
\langle Z\rangle&=  N\int d^N\psi(1)d^N\psi(2) e^{\sum_i \psi_i(1)\psi_i(2)}\int dg \frac{d\sigma}{2\pi i}e^{\sigma[\sum_i \psi_i(1)\psi_i(2)-N g] + \frac{J^2}{q}g^q}\\&= N\int dg \frac{d\sigma}{2\pi i} e^{N[\log(1 + \sigma)  - \sigma g + \frac{J^2}{q}g^q]}.
\end{align}
We would now like to describe how to make sense of this integral. The defining contour of integration for $\sigma$ is along the imaginary axis, and we start by formally integrating $g$ along th real axis. We then evaluate the integral as follows: if we bring down the $\log(1+\sigma)$ term and expand in powers of $\sigma$, we will have $\sigma$ integrals of the form $\frac{N}{2\pi i}\int d\sigma \sigma^p e^{-N\sigma g} = N^{-p}(-\partial_g)^p\delta(g).$ This leads to
\begin{align}\label{firstLinetwopts}
\langle Z\rangle &= \int dg e^{N\frac{J^2}{q}g^q}(1-N^{-1}\partial_g)^N \delta(g) = (1 + N^{-1}\partial_g)^Ne^{N\frac{J^2}{q}g^q}\Big|_{g = 0} \\
&=\sum_{m = 0}^{\lfloor N/q \rfloor} \frac{N!}{(N - m q)! m!}\left(\frac{J^2}{N^{q-1}q}\right)^m.\label{ansTwoPts}
\end{align}
This is the right answer, and we got it from an integral over bosonic variables, but the final $g$ integral was supported in a neighborhood of the origin, and the calculation reduced rather trivially to a direct fermionic computation of (\ref{starthere}).

However, we can also change the contour and make the integral more manifestly well-defined. We rotate the $g$ and $\sigma$ contours in opposite directions by $e^{i\pi/q}$. Here it is simplest to define new variables $\tilde{\sigma} = -i\sigma e^{-i\pi/q}$ and $\tilde{g} = e^{i\pi/q}g$. Then we have
\be
\langle Z\rangle = N\int d\tilde{g}\frac{d\tilde{\sigma}}{2\pi}e^{N[\log(1 + ie^{i\pi/q}\tilde{\sigma}) - i\tilde{\sigma}\tilde{g} - \frac{J^2}{q}\tilde{g}^q]},
\ee
where we integrate $\tilde{g},\tilde{\sigma}$ over the real axis. It is easy to check that numerical integration first over $\tilde{g}$ and then over $\tilde{\sigma}$ indeed gives the correct answer (\ref{ansTwoPts}) for a few values of $N,q$.

One can also discuss saddle points for this integral. For these purposes we go back to the $g,\sigma$ variables. There are $q$ saddle points, the solutions of the equations
\be
\sigma = J^2 g^{q-1}, \hspace{20pt} g = \frac{1}{1+\sigma}.
\ee
There is one real solution, and this is the one that naively dominates. We have not analyzed the deformation of the integration contour in detail, but we observe that this leading saddle does in fact give the right large $N$ behavior, comparing to (\ref{ansTwoPts}).

A confusing aspect of the $G,\Sigma$ representation is that the fundamental variables are Grassmann variables, and we could ask how this is consistent with a representation by $g,\sigma$. For example, the fact that the square of a Grassmann vanishes should imply that $g^{N+1} = 0$. This seems inconsistent with the fact that we are integrating over nonzero values of $\tilde{g}$, and indeed studying saddle points with $g$ nonvanishing. In fact, one can check that an insertion of $g^p$ with $p>N$ will make the integral zero. This is easiest to see from (\ref{firstLinetwopts}), based on the fact that we are at most taking $N$ derivatives of the integrand before setting $g = 0$, so a term of degree $N+1$ will give zero.

\section{Subleading saddle points in the $G,\Sigma$ variables}
\label{subleadingsaddles}
Besides the standard saddle point that gives the themodynamics discussed in section \ref{thermo}, there are a family of subleading saddles for the path integral (\ref{SGS}). We do not have their explicit form for $q= 4$, but we can understand some of their properties numerically, and by comparison to the simpler $q = 2$ theory.

In the $q = 2$ model, the saddle point equations for different Matsubara frequencies decouple, and we have
\be
\label{q2seqns}
G(\omega_n)^{-1} = -i\omega_n - \Sigma(\omega_n), \hspace{20pt} \Sigma(\omega_n) = J^2G(\omega_n)
\ee
with solutions
\be
\label{q2ssolns}
G_\pm(\omega_n) = \frac{-i\omega_n \pm i\text{sgn}(\omega_n)\sqrt{\omega_n^2 + 4 J^2}}{2J^2}.
\ee
Choosing $G_+$ for all frequencies gives the dominant saddle. Choosing $G_-$ for some of the frequencies will lead to subdominant saddles. The difference in saddle point action induced by choosing $G_-$ (for both $\omega_n = \frac{2\pi}{\beta}(n+1/2)>0$ and the corresponding $-\omega_n$) is
\begin{align}
-I(G_+) + I(G_-) &= N\log\frac{1 + \sqrt{4J^2/\omega_n^2+1}}{1 - \sqrt{4J^2/\omega_n^2+1}} +N \frac{|\omega_n|\sqrt{4J^2+\omega_n^2}}{2J^2} \\&= N\left(i\pi + \frac{4\pi(n+\frac{1}{2})}{\beta J} + O(\frac{1}{(\beta J)^3})\right).
\end{align}
For large $\beta J$, we see that the saddles become almost degenerate. Naively, this would suggest a soft mode connecting the saddles, but because the imaginary part differs by an order one amount, we do not have such a mode. However, at large $\beta J \gg N$ one would have to sum over all of these saddles. We will see that they play an important role in appendix \ref{qequal2}.

\begin{figure}
\begin{center}
\includegraphics[width = 0.48\textwidth]{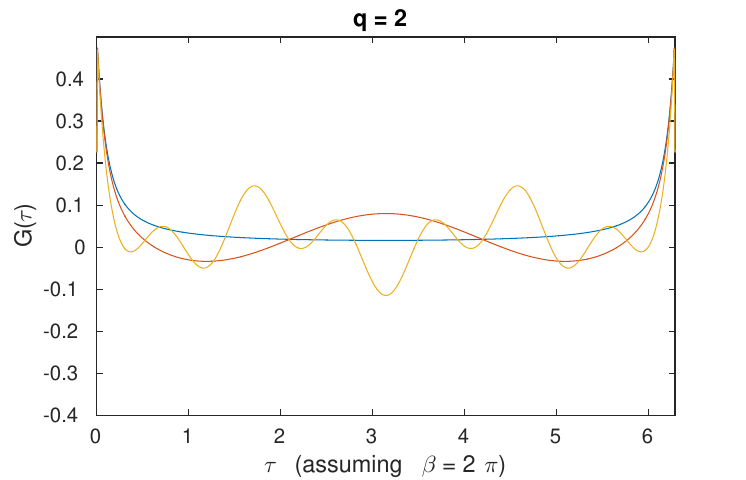}
\includegraphics[width = 0.48\textwidth]{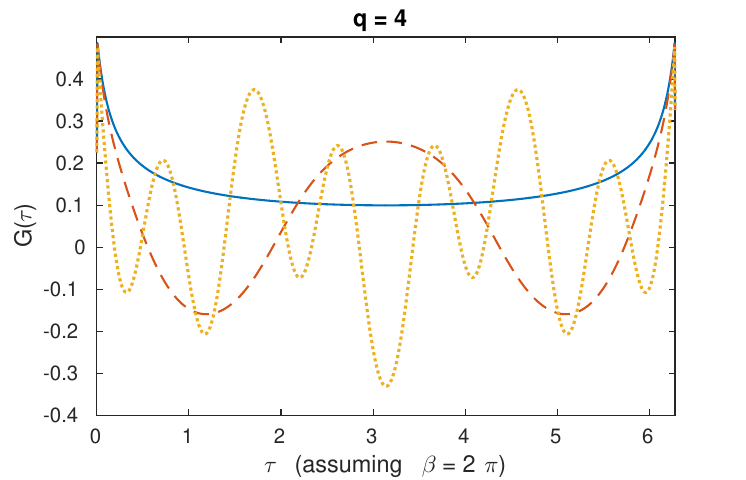}
\caption{At left we show the $q = 2$ standard saddle (blue/solid) and a subleading saddle with $n = 1$ flipped (red/dashed) and one with both $n = 2,6$ flipped (orange/dotted). At right we have the corresponding $q =4$ solutions. We use $\beta\mathcal{J} = 20\pi$.}\label{saddleFig}
\end{center}
\end{figure}
In the $q = 4$ model we do not have explicit formulas, but we can find subleading saddles numerically. It seems that for each $q = 2$ solution there is a corresponding $q = 4$ solution, which can be found by starting with the $q = 2$ solution and iterating the Schwinger-Dyson equations while slowly increasing $q$ from two to four. We give a plot of some solutions in Fig.~\ref{saddleFig}. An important difference between the $q=2$ and $q=4$ cases is that the actions do not become degenerate at large $\beta J$. For the simplest case, where we start with a $q = 2$ solution with a single frequency pair $\omega_n$ flipped, we find numerically that the $q = 4$ action is given by
\be
-I(G_{standard}) + I(G_{subleading}) \approx N\left(i\pi + \frac{n+1}{2}+O(\frac{1}{\beta J}) \right).
\ee
We are not sure that this simple expression is exactly correct, only that it is within a percent or so of the numerical answer for the first few frequencies $n = 0,...,5$ where we were able to check. The important point is that there is a large $\frac{N}{2}$ gap in the action even at very low temperature. This explains why these additional saddles do not disturb the large $N$ thermodynamics. A logical possibility is that the relative dominance of these saddles could change when we study complex $\beta$, but preliminary investigation suggests that this is not the case, and that the gap remains.

Finally, we will mention that in the $q = 4$ theory there also appear to be saddle points that depend nontrivially on both of the time arguments, not just the difference. In other words, we have saddle points that spontaneously break time translation invariance. We have not studied these systematically, but the examples that we found numerically had larger action than the standard saddle.

\section{Saddle points and the $q=2$ model}
\label{qequal2}

 The $q=2$ model is qualitatively different than the model with $q> 2$, since it is equivalent to a model of free fermions with a random mass matrix \cite{Magan:2015yoa, Anninos:2016szt, Gross:2016kjj}. It is not a chaotic system, but the explicit $N\times N$ random matrix leads to a ``mini-ramp'' and ``mini-plateau'' in certain quantities, with plateau time $t_p\sim N$ instead of $t_p\sim L$. In this appendix we show how the saddle points discussed in Appendix \ref{subleadingsaddles} contribute to this behavior.
    
The Hamiltonian of the $q=2$ model is
    \be
       H=i\sum_{i<j}J_{ij}\psi_i\psi_j,
     \ee 
 where $J_{ij}$ is a real antisymmetric matrix. Conjugating with an orthogonal matrix $Q$, we can take $J_{ij}$ to a block diagonal form with each block given by
   \be
    \begin{pmatrix} 0 &\lambda_k \cr -\lambda_k & 0 \end{pmatrix}
 \ee
where $\lambda_k>0$ and with $k$ running from $1$ to $N/2$.
    Then the Hamiltonian can be written as
    \be
    H=i\sum_{k=1}^{N/2} \lambda_k \tilde{\psi}_{2k-1}\tilde{\psi}_{2k} = \sum_{k=1}^{N/2} \lambda_k (c^\dag_k c_k -\frac{1}{2}).
   \ee
    Where $\tilde{\psi}_i=(Q\psi)_i$, and we made Dirac fermions out of these pairs of Majoranas, $c_k=(\tilde{\psi}_{2k-1}+i \tilde{\psi}_{2k})/\sqrt{2}$ and $c^\dag_k=(\tilde{\psi}_{2k-1}-i \tilde{\psi}_{2k})/\sqrt{2}$. Notice that $iJ_{ij}$ is a skew Hermitian matrix, not a GUE Hermitian matrix.   Its eigenvalue statistics are known \cite{mehta2004random, Gross:2016kjj}.  At large $N$ the spectrum is a semicircle, with $1/N$ corrections.  Eigenvalue (mass) pair correlations $R_2(\lambda, \lambda')$ are described by a modified sine kernel whose short distance behavior is that of GUE.      
 
    It follows that eigenvalues in the single particle sector will repel, because of the usual eigenvalue repulsion of a random matrix. However, nearby multiparticle eigenvalues coming from sectors with very different particle numbers will repel only weakly. Because the eigenvalues that repel each other have an average spacing $\sim 1/N$ instead of $1/L$, we expect that the plateau time in this model is $t_p\sim N$.
    
    The simplest observable in this model with a ramp is the (quenched) disorder averaged squared correlation function. It turns out this is easier to calculate than $g_d(t)$.  The averaged correlation function (not squared) does not have a ramp. 
    
    Part of the reason that the correlation functions are easier to calculate is that the matrix elements of $\psi_i$, $\big<n|\psi_i |m\big>$, are only nonzero for $| n \big>$, $|m \big>$ belonging to particle number sectors differing by a particle number of one. This means that the correlation function only receives contributions from energy differences that are equal to the single particle sector energies, making the calculation much simpler. Explicitly, the Euclidean quenched correlation function is
    
    \be
    G(\tau)= \frac{1}{N}\sum_{i=1}^N \bigg<\frac{\text{Tr} [e^{-\beta H} \psi_i(\tau)\psi_i]}{Z(\beta)}\bigg>_J=\int d\lambda \frac{1}{e^{-\beta \lambda}+1} \tilde{\rho}(\lambda) e^{-\lambda\tau}
    \ee
    Here $\tilde{\rho}(\lambda)$ is the average mass density. In the above integral we are extending it to a symmetric function $\tilde{\rho}(-\lambda) = \tilde{\rho}(\lambda)$. We can see that the real time correlation function, obtained by continuing $\tau\rightarrow i t$ in the above expression, will not have a ramp or plateau. However, the connected part of the quenched disorder averaged square of the correlation function will have a ramp and plateau
    
    \be \label{gcsquared}
    G^2_c(\tau,\tau')=\frac{1}{N^2}\sum_{i,j=1}^N \bigg<\frac{\text{Tr}[e^{-\beta H} \psi_i(\tau)\psi_i]\text{Tr} [e^{-\beta H} \psi_j(-\tau')\psi_j]}{Z(\beta)^2}\bigg>_J- G(\tau)G(-\tau')
    \ee
    \be
    =\int d\lambda d\lambda' \frac{1}{e^{-\beta \lambda}+1} \frac{1}{e^{-\beta \lambda}+1} R_2(\lambda,\lambda') e^{-\lambda_1\tau + \lambda_2 \tau'}
    \ee
    Note that the annealed correlator cannot be written simply in terms of $R_2(\lambda,\lambda')$. After analytically continuing $\tau\rightarrow it $ and $\tau'\rightarrow it'$, because of the presence of $R_2(\lambda,\lambda)'$, $G^2_c(t,t')$ will have a ramp and plateau. In particular, at $\beta=0$ it is precisely equal to $g_c(t)$ for the ensemble of skew Hermitian matrices.

    This simple expression for the square of the averaged correlation function in terms of the mass pair correlator suggests that it may be possible to calculate in a simple way by saddle point. Kamenev and Mezard \cite{kamenevmezard} calculated $R_2(\lambda,\lambda')$ in the GUE by saddle point with an integral that is very similar to the path integral \eqref{gsigint} with $q=2$.\footnote{The integral that \cite{kamenevmezard} calculated is closer to the integral \eqref{gsigint} over only one of the Matsubara frequency modes}  This  is why we want to consider the quenched disorder averaged correlation function instead of the annealed disorder average correlation function (where we would J average the denominator and numerator in \eqref{gcsquared} separately).

    The quenched disorder averaged squared correlation function in Matsubara frequency space, $G^2_c(\omega_n,\omega_m)$, can be calculated by coupling sources $z_n$ to the operators $\sum_{i=1}^N \psi_i(-\omega_n)\psi_i(\omega_n)$ with a term in the action $S\supset \frac{1}{2}\sum_{i=1}^N\sum_{n=0}^\infty \psi_i(-\omega_n)\psi_i(\omega_n) z_n$. Let $Z(\{z\})$ be the partition function with the source term included,
    
    \be
    G^2_c(\omega_n,\omega_m)= \frac{1}{N^2} \frac{\partial}{\partial z_n }\frac{\partial}{\partial \tilde{z}_m} \bigg(\big<\log Z(\{z\}) \log Z(\{\tilde{z}\})\big>_J-\big<\log Z(\{z\})\big>_J\big< \log Z(\{\tilde{z}\})\big>_J\bigg)\bigg|_{\{z\},\{\tilde{z}\}=0}
    \ee
    The key simplification is that the partition function is a product over all the frequencies, $Z(\{z\})=\prod_{n=0}^\infty Z_n (\{z_n\})$, and since the logarithms of these products turn into sums over the different frequencies, the derivatives simplify. We find
    
    \be
    G^2_c(\omega_n,\omega_m)=\frac{1}{N^2} \frac{\partial}{\partial z_n}\frac{\partial}{\partial \tilde{z}_m} \bigg(\big<\log Z_n(z_n) \log Z_m(\tilde{z}_m)\big>_J-\big<\log Z_n(z_n)\big>_J\big< \log Z_m(\tilde{z}_m)\big>_J\bigg)\bigg|_{z_n,\tilde{z}_m=0}
    \ee
    
    Now we evaluate the averaged logarithms of the single frequency factors of the sourced partition function. This is almost exactly the calculation of Kamenev and Mezard \cite{kamenevmezard}.\footnote{Their calculation applied to the GUE, while ours applies to the ensemble of skew Hermitian matrices. The difference between our integrals comes from the reality constraint on the Majoranas, which gives a different result at order 1/N.} They use the replica trick to rewrite the average of the logarithm in terms of the average of the replicated partition function. They then evaluate the replicated partition function by the saddle point approximation. Their saddle point equation for the average of a single logarithm is equivalent to the equation obtained by combining the saddle point equations for $G(\omega_n)$ and $\Sigma(\omega_n)$ for one frequency \eqref{q2seqns}, except that we now account for the source with a shift of $\omega_n$, $-i\omega_n\rightarrow-i\omega_n+z_n$.

     For the average of the product of logarithms, the saddle point equations have a mixing term. These equations are quadratic and thus have two solutions, $G_+(\omega_n)$ and $G_-(\omega_n)$ \eqref{q2ssolns}. Choosing a replica symmetric solution with $G_+$ for each replica gives the dominant contribution to the integrals, the fluctuation is the first term that survives. These contributions correspond to the semicircle part of the mass distribution and mass pair correlation function, and the fluctuations give the ramp. Considering a replica symmetry breaking saddle point involving both $G_+(\omega_n)$ and $G_-(\omega_n)$ gives the sine kernel type contribution to $R_2(\lambda,\lambda')$ in $G_c^2(\omega_n,\omega_m)$, and thus gives us the plateau.
     
As we noted above, calculations of $Z(t)$ and $g(t)$ using saddle points will be more complicated. It appears that the Itzykson-Zuber integral \cite{Itzykson:1979fi} will be helpful.  We hope to return to this issue in future work.

\section{On $N^{-q}$ vs. $2^{-N}$}
\label{Nq}
It would be nice to have a direct analytical argument for the ramp and plateau in SYK. As a first step, one would like to understand where the $e^{-2S}$ scale of the ramp comes from. Naively, this is puzzling, because the ramp arises from correlations between the two replicas, and in simple diagrams such correlations are suppressed by powers of $N^q$, not exponential factors. In this appendix, we make a simple comment about how the exponential can emerge from such diagrams.

We start by defining the quantity
\be\label{Fk1k2}
F_{k_1,k_2} \equiv \frac{\langle \tr H^{k_1} \tr H^{k_2}\rangle}{L^2\sigma^{k1+k2}}, \hspace{20pt} \sigma^2 = \frac{1}{L}\langle \tr H^2\rangle.
\ee
In principle, knowing $F_{k_1,k_2}$ makes it possible to evaluate the double resolvent $\langle \tr\frac{1}{z-H}\tr\frac{1}{w-H}\rangle$. By taking discontinuities in both $z$ and $w$ across the real axis, one gets an expression for the pair correlation function $\langle \rho(z)\rho(w)\rangle$, which gives rise to the ramp.

This procedure has been carried out for the GUE ensemble by Brezin and Zee \cite{Brezin:1993qg}. At leading order in $1/L^2$, one considers planar graphs only: most of the Wick contractions do not contribute, and many of the remaining graphs for the double resolvent can be summed by replacing $z,w$ by dressed propagators. All that remains is a special class of graphs where we take $k_1 = k_2 = k$ and then Wick-pair the Hamiltonians in (\ref{Fk1k2}) ``straight across'' up to an overall reflection. More explicitly, the first $H$ factor in the first trace is paired with the $k$-th factor in the second trace. The second factor in the first trace is paired with the $(k-1)$-st factor in the second trace, and so on. We refer to the result of this special contraction as $f_k$.\footnote{The contribution of all contractions related to such a configuration by cyclicity would be $k f_k$.} In GUE one finds $f_k = 2^{-N}$, which is the origin of the $2^{-N}$ coefficient of the ramp. The linear time dependence arises from a singularity in the geometric series that defines the double resolvent, and in particular is sensitive only to the $f_k$ for large $k$. This is an important point so we will emphasize it: the short-distance correlations between eigenvalues, or equivalently the late-time behavior of the ramp, is related to the large $k$ behavior of the $f_k$ or $F_{k_1,k_2}$ coefficients.

In SYK, the class of graphs that must be summed at leading order is larger than in GUE. In particular, we have to think about the $1/N$ expansion instead of the $1/L$ expansion. We will not attempt to analyze the sum in a systematic way. Instead, we will simply comment on the behavior of the special class of graphs that we used to define $f_k$ above, because these already provide a model for the handoff between $N^{-q}$ and $2^{-N}$.

We define $f_k$ the same way as above: we let $k_1 = k_2$ in (\ref{Fk1k2}) and we Wick-contract the couplings in each factor of $H$ in the first trace with the corrresponding (reflected, as before) factor of $H$ in the second trace. This is equivalent to the following: we imagine writing a product of $k$ of the possible terms that appear in the Hamiltonian. Then $f_k$ is simply the probability that such a product has a nonzero trace. For small values of $k$, $f_k$ is suppressed by powers of $N^q$, as expected for a two-replica correlation. For example, $f_2 = \frac{1}{{N\choose q}}\sim N^{-q}$. However, for large values of $k$, $f_k$ approaches a constant value of $2^{1-N}$. This is because for a product of fermions to have a nonzero trace, we must have an even number of each flavor of fermion, leading to $N$ binary constraints. The exact formula is
\begin{align}
f_k &= 2^{-N}\sum_{\{x_i = \pm 1\}}\left( \sum_{i_1<...<i_q}x_{i_1}...x_{i_q}\right)^k= 2^{-N}\sum_{m = 0}^N {N\choose m}\left(\sum_{p = 0}^q {m\choose p}{N-m\choose q-p}(-1)^p\right)^k\notag\\
&= 2^{-N}\sum_{m = 0}^N {N\choose m}\alpha(N,m,q)^k, \hspace{20pt} \alpha \equiv \frac{(N{-}q)!(N{-}m)!}{N!}\frac{{}_2F_1(-m,-q,N{-}m{-}q{+}1,-1)}{\Gamma(N{-}m{-}q{+}1)}.\notag
\end{align}
For large values of $k$ the largest $\alpha$ dominates the sum. This is the value $\alpha=1$ when $m = 0,N$, which leads to $f_k \approx 2^{1-N}$. In particular, for the large values of $k$ relevant for the late-time ramp, we find the same behavior as in GUE, for any value of $q$. We suspect that this is a hint of the universality of local random matrix statistics, and is the basic point behind the origin of the $e^{-2S}\sim 2^{-N}$ ramp. 

We are currently working to make this more precise.

\section{Constraints on saddle point origins of the ramp}
\label{Gramp}

As explained in Section~\ref{rampRMT}, the late time plateau is a highly non-perturbative effect in SYK that is expected to involve effects  as small as  $\exp(-e^N)$,  based on random matrix theory analysis.
On the other hand, the ramp scales as $e^{-N}$ and so it may be a more tractable non-perturbative effect.
In particular, random matrix theory tells us that the part of the ramp that is linear in time is a perturbative effect in RMT, and this part may be an ordinary non-perturbative effect in SYK.

In this appendix we make a few comments about the  simplest possible approach to explaining the ramp --  finding a nontrivial saddle of the original $G, \Sigma$ action.  But because $G$ is small the source $\log G$ in the action will deform the saddle point.  There is backreaction.

Such a saddle would have to satisfy  constraints.  First, in order to account for the N mod 8 periodicity discussed in Section \ref{sec:corr} there would have to multiple saddles with complex action. 

  The second constraint is more nontrivial.  As discussed in Section~\ref{single}, the ramp and plateau are not self-averaging (both in the two-point function and in the spectral form factor) \cite{prange1997spectral}.
The fluctuations on the ramp are of the same size as its mean value.
But a saddle point explanation requires that we have a limit in which fluctuations are suppressed.  

This argument may seem a bit quick because the large fluctuations we are discussing are in the integral over random couplings, but this integral can be performed exactly.
In particular, in the $G,\Sigma$ formulation the disorder integral is done first, followed by the integral over the fermion variables, and we are left with an integral over the $G,\Sigma$ variables.
We checked that the latter integral also exhibits large fluctuations on the ramp and plateau (of the same order as the mean value), by numerically comparing the variance $\langle G(t)^2 \rangle - \langle G(t) \rangle^2$ with the mean, directly in the original fermion formulation.

It is possible that the saddle point backreaction for $\langle G \rangle$ and for $\langle G^2 \rangle$ is delicately tuned to make these answers consistent with numerics, but we see no obvious mechanism for this.

\section{Data}
\label{dataapp}
This section contains some further numerical results. We first present $g(t)$, $g_c(t)$, and $g_d(t)$ for $\beta=0, 1, 5$  for $N = 16, 18, \ldots, 34$ and discuss the dip-ramp-plateau features of $g$ and $g_c$, which exhibit the mod-8 symmetry pattern.
The methods for determining the dip time $t_d$ and the plateau time $t_p$ are explained next, with the results for $N = 10, 12, \ldots, 34$.  We compare the fit of $t_d$ with an exponential and a power-law function.   The error bars are large but the results for larger $N$ are consistent with the estimate in Section \ref{rampSYK}.  They are also consistent with other scenarios involving a crash at earlier time.   The available $N$ values are not large enough to disentangle all these effects.

The plateau time $t_p$ shows a faster exponential increase, and the numerical result is compared with the results of Sections \ref{thermo} and \ref{rampSYK}.
This, together with the results for $t_d$, show that the ramp length grows exponentially in $N$. For $N=34$ we have fitted the ramp power law omitting times near $t_p$ where unfolding effects are important. We find a power consistent with the GUE behavior $g(t) \sim t^{1}$ within a couple percent.

All $g(t)$, $g_c(t)$, and $g_d(t)$  data discussed so far has been for factorized (annealed) quantities, as in \eqref{g}--\eqref{gc}. We compare with the results of the unfactorized (quenched) versions in Sec.~\ref{App:f-uf}.

Finally, in \ref{App:rho} we plot the average density of states for different values of $N$.

\subsection{Plots of $g(t)$, $g_c(t)$, and $g_d(t)$}
\label{App:ggcgd}

\begin{figure}[h]
	\centering
	\includegraphics[width=0.8\textwidth]{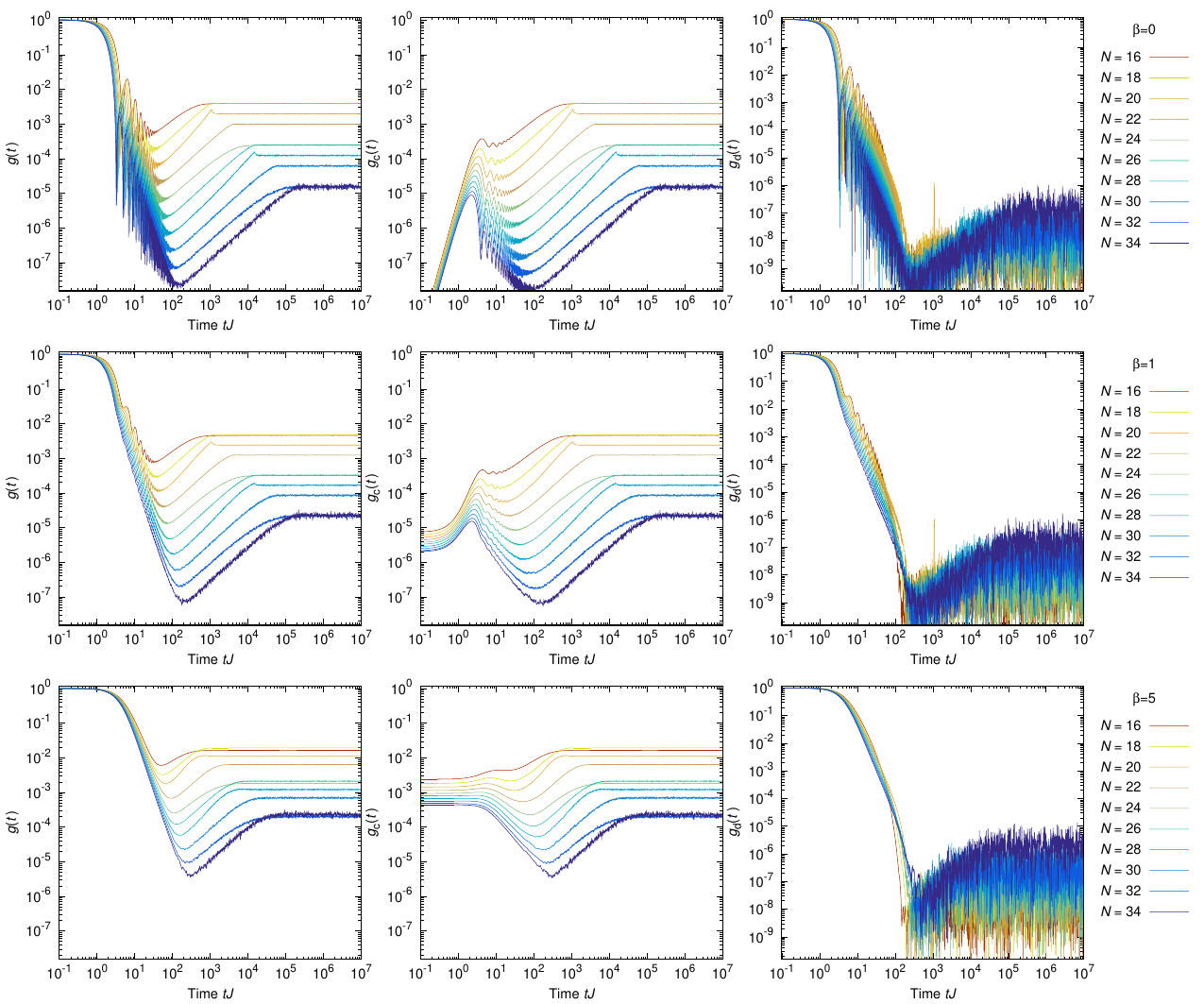}
	\caption{Plots of $g(t)$, $g_c(t)$, and $g_d(t)$ for $N=16, 18, \ldots, 34$ and $\beta = 0, 1, 5$, from top to bottom. The noisy part of the curves for $g_d$ are due to the finite number of samples. We expect the true disorder average to continue decreasing rapidly.}
	\label{fig:f-ggcgdt}
\end{figure}
In Fig.~\ref{fig:f-ggcgdt} we plot $g(t)$, $g_c(t)$, and $g_d(t)$ on a  log-log scale. The oscillation observed for $\beta=0$ before the dip time is also visible for $\beta=1$ but becomes negligible for $\beta=5$.  It is due to interference between the upper and lower edges of the eigenvalue distribution.

$g_d(t)$ decays quickly to typically much smaller values than $g(t)$ or $g_c(t)$ around the dip time.   This is consistent with the theoretical expectation of a gaussian falloff due to fluctuations in the edge of the eigenvalue distribution at times of order $N$ (albeit with a somewhat large coefficient).   Such effects cancel out in $g(t)$.
(Beyond the dip time $g_d(t)$ seems to rebound. This is just because the number of samples is finite and hence the cancellation is not perfect.)  

Around the plateau time, the curves for $g(t)$ and $g_c(t)$ exhibits a sharp peak for $N=20$ and $28$ (GSE), a kink for $N=18, 22, \ldots, 34$ (GUE), and a smoother connection for $N=16, 24, 32$ (GOE),
for $\beta=0$. For $\beta=1$ the feature is preserved, while for $\beta=5$ the peak is broadened and the kink is less visible. However, the plateau heights for $N\not\equiv 0\ (\mathrm{mod}~8)$ (GUE and GSE) cases appear shifted up compared to those for $N\equiv 0\ (\mathrm{mod}~8)$ (GOE) cases, for all values of $\beta$, and the plateau heights for $N=18, 26, 34$ are higher than those for $N=16, 24, 32$ for $\beta>0$.   All of this is consistent with the RMT interpretation, symmetry considerations  and smoothing due to unfolding effects.

\subsection{Dip time $t_d$, plateau time $t_p$ and plateau height}
\label{App:dtime}

Intuitively, the dip time can be determined by finding the minimum value of $g$. 
However, with finite statistics, the error is large because of the non-self-averaging nature of $g(t)$ past the dip. 
Therefore, we estimated the error bar as follows. Firstly we found the minimum value $g_{\rm min}$. 
Then, the lower and upper limits of the error bar are estimated as the smallest and largest $t$ 
which give $g(t)<g_{\rm min}\times 1.04$. 

We can fit $t_\mathrm{dip}$ with an exponential function of $N$ $t_0\mathrm{e}^{\kappa_d N}$.
 $\kappa_d$ does not exhibit clear dependence on $\beta$ from our data (although we expect a weak dependence theoretically).   The error bars are large but the results for larger $N$ are consistent with the estimate in Section \ref{rampSYK}. A power-law fit ($t_\mathrm{d} \sim t_0' N^{\alpha_d}$) cannot be ruled out from our data up to $N=34$.    Again, the available $N$ values are insufficient for a conclusive analysis here.

\begin{figure}[h]
  \centering
  \includegraphics[width=0.3\textwidth]{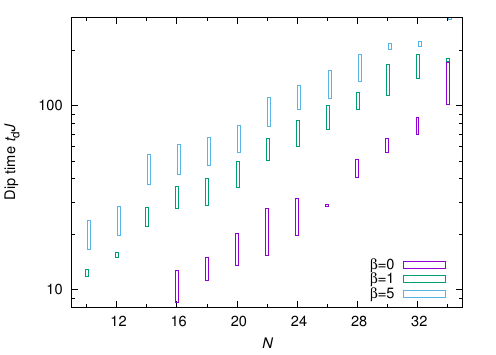}
  \includegraphics[width=0.3\textwidth]{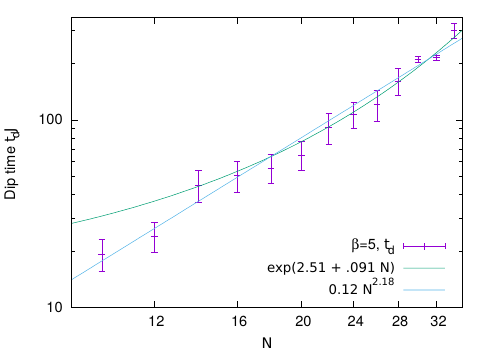}
  \includegraphics[width=0.3\textwidth]{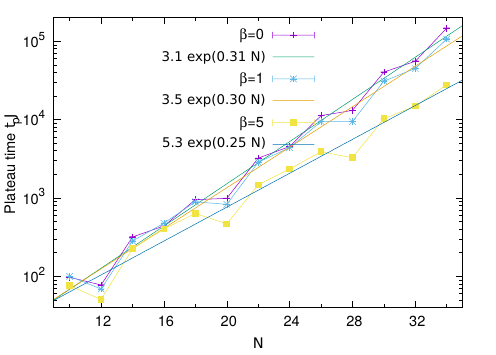}
  \caption{Left: The dip time $t_\mathrm{d}$ against $N$, for $\beta=0,1,5$. The lower and upper limits of the error bar
indicate the range of data points with $g(t) < 1.04~g_\mathrm{min}$.
Middle: Comparison of fits of the SYK $t_\mathrm{d}$ with exponential and power-law functions of $N$ for $\beta = 5$. Right: Plot of the plateau time $t_p$ against $N$, for $\beta = 0,1,5$.
}
  \label{fig:fgtdip-N-mm4SYK_Nm10-34-b5-exppow.pdf}
\end{figure}

As discussed in the main text, the function $g(t)$ reaches a plateau at exponentially late time. Numerically, we find that the height agrees with the expectation $Z(2\beta)/Z(\beta)^2$ when we take an average 
with sufficiently many samples.
The plateau $t_p$ is defined by fitting the ramp by a power-law of the time (linear function in the log-log plot) and the plateau by a constant, and finding the crossing point of the two lines.
We choose the starting point of the fitting range for the ramp 
as $t_s = 5~t_d$ if $g(5~t_d) < 0.7~g_p$, otherwise we use the time at which $g(t_s) = 0.4~g_p$.
The end of the fitting range is the time at which $g(t_e) = 0.7~g_p$, and we fit $\log(g(t))$ by a linear fit and find the time at which the line reaches $\log t_p$. In the right panel of Fig.~\ref{fig:fgtdip-N-mm4SYK_Nm10-34-b5-exppow.pdf} we plot $t_p$ against $N$.

As explained in Section~\ref{rampSYK}, we expect $t_p \sim \mathrm{const.} \exp(S(2\beta))$.
Also, as explained in Section~\ref{thermo}, the expression for entropy at low temperature is
\begin{align}
S(2\beta) = (0.23 + 0.198/\beta+\cdots)N + \cdots.
\end{align}
At $\beta=5$ the coefficient of $N$ is 0.27 up to $O(\beta^{-2})$ corrections, which is close to our numerical result  $0.249 \pm 0.014$. 

As we have seen $t_p\sim \mathrm{e}^{\kappa_p N}$
and $t_d\sim \mathrm{e}^{\kappa_d N}$, where $\kappa_d < \kappa_p = S(2\beta)$.
Hence $\log(t_p/t_d)/N \sim \kappa_p - \kappa_d$ should be constant up to $1/N$.
We observe  that $\kappa_p - \kappa_d > 0$ .
Therefore, the length of the ramp seems to increase exponentially in $N$, consistent with Section \ref{rampSYK}.  Of course our values of $N$ are not large enough to make definitive statements.

\begin{figure}[h]
	\centering
	\includegraphics[width=0.4\textwidth]{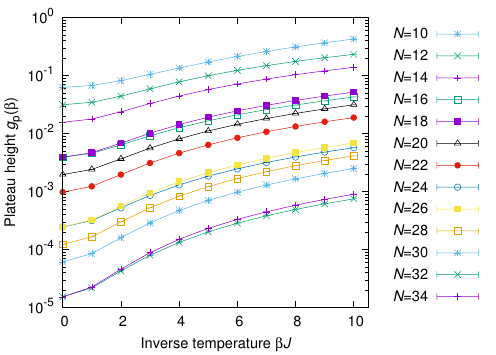}
	\includegraphics[width=0.4\textwidth]{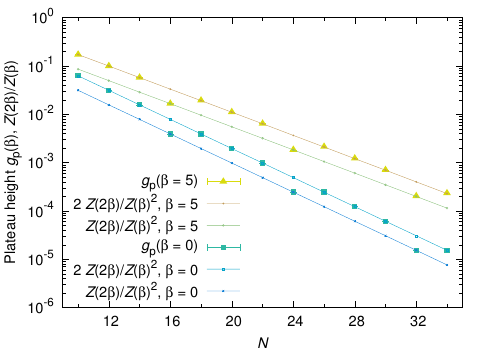}
	\caption{
	Left: Plot of the plateau height against $\beta$ for $N=10,12,\ldots,34$.
	Right: Plot of the plateau height and $Z(2\beta)/Z(\beta)^2$ for $\beta=0,5$ against $N$.
	A clear mod--8 pattern can be seen. For $N\equiv0\ (\mathrm{mod}~8)$, $g_c(\beta)$ equals $Z(2\beta) / Z(\beta)^2$, which for $\beta=0$ equals $1/Z(\beta=0) = 2^{-N/2}$,
otherwise $g_c(\beta) = 2 Z(2\beta) / Z(\beta)^2$ due to the degeneracy in the eigenvalue of the SYK Hamiltonian.}
	\label{fig:f-gt-hp}
\end{figure}
Theoretically the height of plateau of $g(t)$ is $g_p(\beta) = Z(2\beta)/Z(\beta)^2$, \eqref{timeavg}, unless there is degeneracy in the eigenvalues of the model Hamiltonian.
In the SYK model, as has been discussed in Secs.~\ref{nn} and \ref{sec:corr}, all eigenvalues are doubly degenerate when $N~\mathrm{mod}~8 = 2, 4$ or $6$.
Therefore we expect $g_p(\beta) = 2 Z(2\beta)/ Z(\beta)^2$. For $\beta = 0$ this equals the inverse of $Z(\beta = 0) = 2^{N/2}$.
For $N~\mathrm{mod}~8 = 0$, on the other hand, we do not expect eigenvalue degeneracy and thus expect $g_p(\beta) = Z(2\beta)/ Z(\beta)^2$.
We can see nice agreement  in Fig.~\ref{fig:f-gt-hp}.

\subsection{Comparison of factorized and unfactorized quantities}
\label{App:f-uf}

As explained in Section~\ref{spectral}, there are two options for defining the spectral form factor.
Namely, the factorized, or annealed, quantities \eqref{g}, \eqref{gd}, and \eqref{gc}, 
and the unfactorized, or quenched, versions where one averages over $J$ after dividing by $Z(\beta)^2$.
These two choices agree when the quantity of interest is self-averaging (up to order $1/N^{q}$).
Therefore, $g$ and $g_u$ must agree at early time.   Numerically we find they agree at large $N$ for all time.  $g_c(t)$ is not self averaging at early time and so differs from $g_{uc}(t)$ there.

\subsection{Density of states $\rho(E)$}
\label{App:rho}

\begin{figure}[h!]
	\centering
	\includegraphics[width=0.58\textwidth]{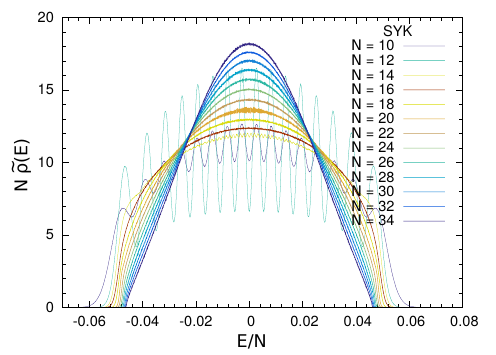}
	\caption{Normalized density of states $\tilde{\rho}(E)$ for the SYK model with $N=10, 12, \ldots, 34$. The bin width is $10^{-3}J$. Notice that the energy is measured in units of $N J$. The numbers of samples are $21 600 000$ ($N=10$), $10 800 000$ ($N=12$), $5 400 000$ ($N=14$), $1 200 000$ ($N=16$),
	600 000 ($N=18$), 240 000 ($N=20$), 120 000 ($N=22$),
	48 000 ($N=24$), 10 000 ($N=26$), 3~000 ($N=28$), 914 ($N=30$), 516 ($N=32$), 90 ($N=34$).
	}
	\label{fig:sp_Nm10-34}
\end{figure}

In Fig.~\ref{fig:sp_Nm10-34} we plot the normalized density of states $\tilde{\rho}(E)$, averaging the spectrum obtained by diagonalizing the Hamiltonian \eqref{Hmajorana} for many disorder parameters.
	Almost periodic oscillations due to level repulsion are clearly observed for small values of $N$. For large $N$ and fixed $q$, the distribution will converge in e.g. an $L^2$ norm sense to a Gaussian \cite{erdHos2014phase}, with width $E\sim \sqrt{N}$. However, the small tails of $\tilde{\rho}$ for energies of order $E\sim N$ will not be described by a Gaussian, and will contain an exponentially large number of states.

\clearpage
\bibliographystyle{utphys}
\bibliography{refs}

\end{document}